\documentclass[english,aps,pra,superscriptaddress,twocolumn,floatfix]{revtex4}
\usepackage[T1]{fontenc}
\usepackage[latin9]{inputenc}
\usepackage{units}
\usepackage{amssymb}
\usepackage{graphicx}
\usepackage[usenames,dvipsnames]{xcolor}
\usepackage{natbib}
\usepackage[caption=false]{subfig}
\usepackage{babel}
\usepackage{array}
\usepackage{multirow}
\usepackage{placeins}
\usepackage{hyperref}
\usepackage{soul}
\usepackage{verbatim}

\usepackage{amsfonts}
\usepackage{bbold}
\usepackage{amsmath}
\usepackage{siunitx}
\usepackage{epsfig}

\newcommand{\Identity}{\mathbb{1}}

\newcommand{\ket}[1]{
  \vert #1  \rangle
}
\newcommand{\bra}[1]{
  \langle #1 \vert
}

\begin{document}

\title
{A silicon-based surface code quantum computer}


\author{Joe O'Gorman}
\address{Department of Materials, University of Oxford, Parks Road, Oxford OX1 3PH, UK}
\author{Naomi H. Nickerson}
\address{Department of Physics, Imperial College London, SW7 2AZ, UK}
\address{Department of Materials, University of Oxford, Parks Road, Oxford OX1 3PH, UK}
\author{ Philipp Ross}
\address{London Centre for Nanotechnology, 17-19 Gordon St., London WC1H 0AH }
\author{John J. L. Morton\footnote{Address correspondence to jjl.morton@ucl.ac.uk}}
\address{London Centre for Nanotechnology, 17-19 Gordon St., London WC1H 0AH }
\address{Department of Electronic and Electrical Engineering, University College London, Torrington Place, London WC1E 7JE}
\author{Simon C. Benjamin\footnote{Address correspondence to s.benjamin@qubit.org}}
\address{Department of Materials, University of Oxford, Parks Road, Oxford OX1 3PH, UK}

\begin{abstract}
Individual impurity atoms in silicon can make superb individual qubits, but it remains an immense challenge to build a multi-qubit processor: There is a basic conflict between nanometre separation desired for qubit-qubit interactions, and the much larger scales that would enable control and addressing in a manufacturable and fault tolerant architecture. Here we resolve this conflict by establishing the feasibility of surface code quantum computing using solid state spins, or `data qubits', that are widely separated from one another. We employ a second set of `probe' spins which are mechanically separate from the data qubits and move in-and-out of their proximity. The spin dipole-dipole interactions give rise to phase shifts; measuring a probe's total phase reveals the collective parity of the data qubits along the probe's path. We introduce a protocol to balance the systematic errors due to the spins being imperfectly located during device fabrication. Detailed simulations show that the surface code's threshold then corresponds to misalignments that are substantial on the scale of the array, indicating that it is very robust. We conclude that this simple `orbital probe' architecture overcomes many of the difficulties facing solid state quantum computing, while minimising the complexity and offering qubit densities that are several orders of magnitude greater than other systems. 

The code written for our numerical simulations is openly available online~\cite{ourCode}.
\end{abstract}
\maketitle

The problem of scalability remains one of the great challenges facing the development of quantum computers. For {\em classical} information processing, the semiconductor revolution enabled a spectacularly successful scaling that has led to today's highly complex consumer devices. It is reasonable to hope that some of this vast expertise could be fruitfully brought to bear on quantum systems. An influential early paper exploring this possibility was written by Kane~\cite{kane98} in 1998. According to this proposal, impurity atoms implanted in a pure silicon matrix constitute the means of storing qubits. Operations between qubits would occur through {\em direct} contact interactions between such spins, which necessitated an inter-qubit spacing of at most nanometers (and therefore a precision considerably greater than this) together with exquisitely small and precisely aligned electrode gates to modulate the interaction. This proposal proved highly influential and progress toward realising it has been made both through theoretical work advancing the architecture~\cite{Hollenberg2006} and at the experimental level, including impurity positioning via STM techniques that have achieved nanometer precision~\cite{Schofield2003, Fuechsle2012}. However it remains extremely challenging as a path to practical quantum computing. 

Since 1998 there has been dramatic progress in understanding the representation and processing of quantum information. Surface codes have emerged as an elegant and practical method for representing information in a quantum computer. The units of information, or logical qubits, can be encoded into a simple 2D array of physical qubits~\cite{topolo}.  By measuring stabilizers, which essentially means finding the parity of nearby groups of physical qubits, errors can be detected as they arise. Moreover with a suitable choice of stabilizer measurements the encoded qubits can even be manipulated to perform logical operations. The act of measuring stabilizers over the array thus constitutes a fundamental repeating cycle for the computer and all higher functions can be built upon it. Importantly, all the required parity measurement operations can be made locally within a simple 2D array, and various studies have established a high level of fault tolerance -- of order $1\%$ in terms of the probability for a low level error in preparation, control or measurement of the physical qubits~\cite{FOWLER_2009high,WANG_QCwithNNinteractionsAndErrorRatesOverOnePercent}.

In view of the power and elegance of the surface code picture, one can now revisit the ideas of the Kane proposal and reimagine it as an engine designed `from the bottom up' to efficiently perform stabilizer measurements. This is the task we undertake in the present paper. We find that one can abandon the need for direct gating between physical qubits, and with it the need for extreme precision in the location of impurities and the equally challenging demand for electrical gating of qubit-qubit interactions. This is replaced by a requirement for parity measurement of groups of four spins, which we argue can be performed by a simple repeating cyclical motion. Crucially, we exploit long range dipole fields rather than contact interactions, and we are thus able to select the scale of the device according to our technological abilities. Presently we show that the tolerances in our scheme, i.e. the amounts by which dimensions can be allowed to vary, can be orders of magnitude greater than those demanded in the Kane proposal. A further advantage of our approach is that it requires active control of only the electron spins, rather than the nuclear spins. These various advantages come with a new and unique challenge: the device consists of two {\em mechanically separate} parts, which are continually shifted slightly with respect to one another in a cyclic motion. Deferring a full discussion of practicality to later in the paper, here we simply note that the requirements in terms of the surface flatness and the precision of mechanical control are considerably less demanding than the tolerances achieved in modern hard disk drives.

 \begin{figure}[t]
\includegraphics[width=\columnwidth]{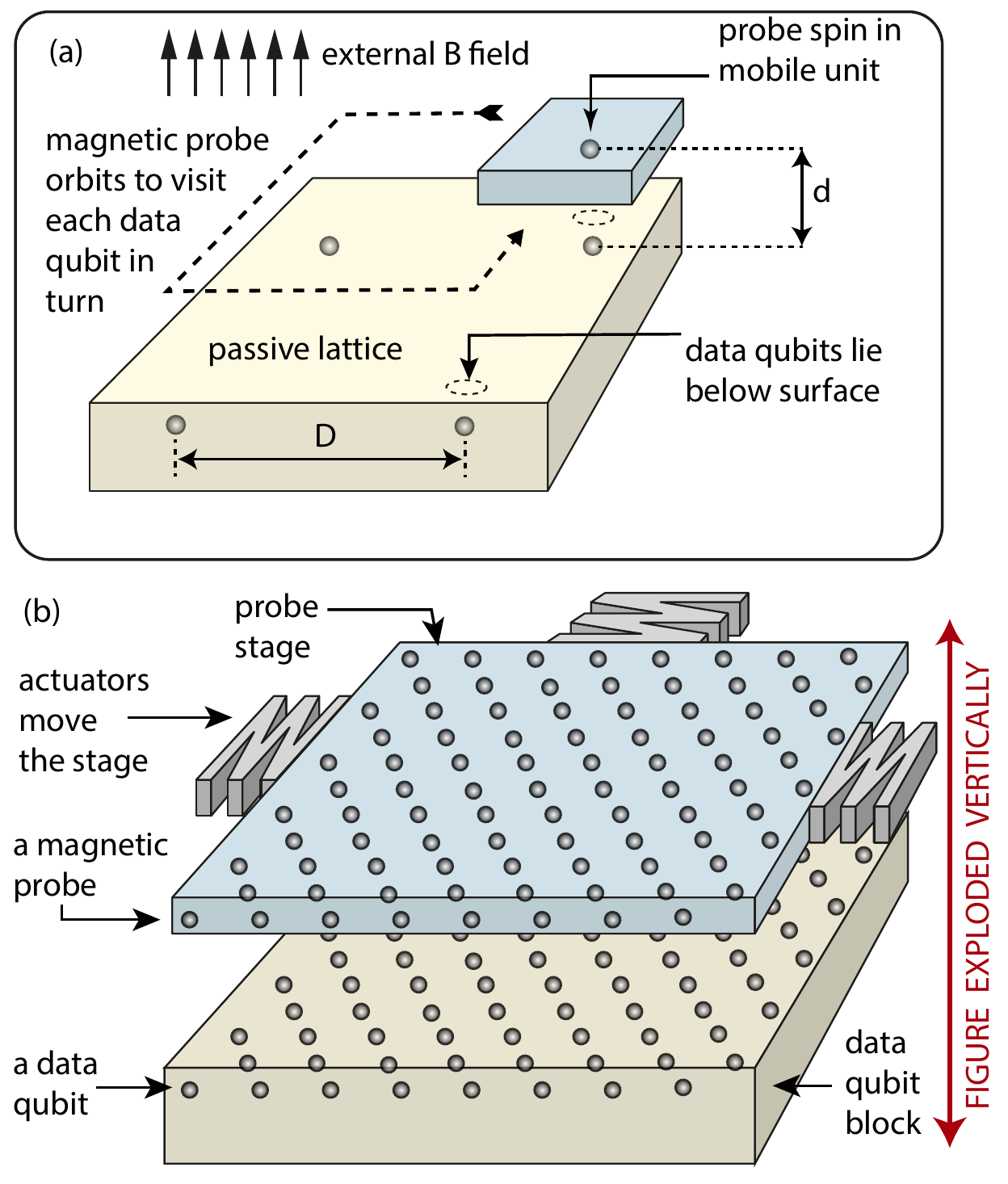}
\caption{\label{basicIdeaFig} (a) The principle of the orbital probe parity measurement: \textcolor{black}{a probe spin comes into proximity with 4 data qubits during one cycle. (b) Simplified schematic of a scalable device showing both that the probe layer and the data qubit layer contain extended spin arrays (details of the their relative positions are shown in Fig.\ref{ProbeLayouts}). Note here the probe stage is shown as mobile while the data qubit stage is static; but in fact either may move, it is their relative motion that is key.}}
\label{architectures}
\end{figure}

We begin with a discussion of the physics of the parity measurement process, before moving on to analyse the robustness of the device against various kinds of imperfection. The essential elements of the scheme are shown in Fig.~\ref{basicIdeaFig}(a). Spin-$\frac{1}{2}$ particles suffice for the protocol we describe, and so we will restrict our analysis to this case, however we do not foresee any basic obstacle to generalising to higher spin systems. In the figure, four spin-$\frac{1}{2}$ particles referred to as `data qubits' are embedded in a static lattice. In practice these are likely to be electrons bound to isolated donor impurities {\color{black} in silicon}, which we describe in more detail later. Meanwhile another spin-$\frac{1}{2}$ particle is associated with a mechanically separate element which can move with respect to the static lattice. We assume this `probe spin' is also electronic, {\color{black} for example, either a different species of donor in silicon or an NV defect centre in diamond.} It will be necessary to prepare and measure the state of the probe; as we presently discuss, this might be achieved  {\color{black} via spin-to-charge conversion for donors in silicon or, alternatively, by optical means for the NV centre.} \textcolor{black}{There are two key dimensions: the vertical distance between a probe qubit and a data qubit at closest approach, $d$, and the  separation between qubits in the horizontal plane, $D$. It is important that $d \ll D$, in order that any interactions between the in-plane spins are relatively weak. As we will discuss, the optimal choice of dimensions varies with several factors including the nature of the mechanical movement and moreover the entire structure can be scaled in proportion; but as an example, for one realisation of the system $d=$ \SI{40}{\nano\meter}  and $D=$ \SI{400}{\nano\meter} will prove to be appropriate. For comparison, note that commercial disk drives can achieve a \SI{3}{\nano\meter} `flying height' between read/write head and platter.} Given this setup, our goal is to measure the parity of the four data qubits i.e. to make a measurement which reports `even' and leaves the data qubits in the subspace  $\{\ket{0000}$, $\ket{1100}$, $\ket{0011}$, $\ket{0110}$, $\ket{1001}$, $\ket{0101}$, $\ket{1010}$, $\ket{1111}\}$, or which reports `odd' and leaves the four data qubits in the complementary subspace.

\begin{figure*}
\centering
\includegraphics[width=2.\columnwidth]{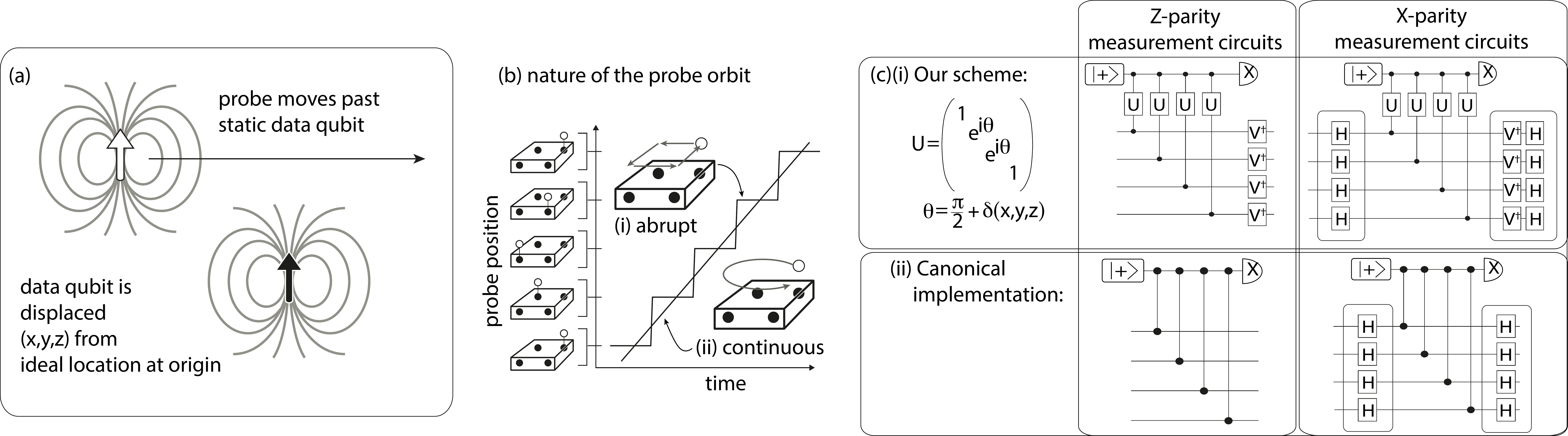}
\caption{\label{interactionFig}The physical process of parity measurement. (a) The probe and a data qubit move past one another and in doing so a state dependent phase shift occurs. (b) We consider two ways in which the probe may move: abruptly, site to site, or in a continuous circular motion. (c) The net phase acquired by a probe as it transits the cycle of four data qubits reveals their parity, but nothing else. \textcolor{black}{(i) Two equivalent circuits for measuring the Z-parity of four data qubits. Using $U=S(\pi/2)$, as we propose and fixing the unconditional phases $V$ on the data qubits is equivalent to (ii) the canonical circuit composed of controlled-phase gates. Note that only global (boxed) operations are required on the data qubits and the X- and Z-parity measurements differ only by global Hadamard operations. A full description of the noisy circuit is deferred to Appendix~\ref{Appendix:AnalyticSO}}.}
\label{circuits}
\end{figure*}
In the abstract language of quantum gates, building a parity measurement is straightforward. The following process is widely used in the quantum computing literature: we prepare an ancilla (the probe, in our case) in state $\ket{+}=(\ket{0}+\ket{1})/\sqrt 2$ and then {perform two-qubit controlled-phase gate $G=diag\{1,1,1,-1\}$ between the probe and one of the four data qubits. We then repeat this operation between the probe and each of the three remaining data qubits in turn. Finally we measure the probe in the basis $\{ \ket{+}, \ket{-} \}$. The quantum circuit for this parity measurement is shown in Fig.~\ref{circuits}(c)(ii)}. If we see outcome $\ket{+}$ then the data qubits are in the `even' space, while $\ket{-}$ indicates `odd'. (This is easy to see by reflecting that the ancilla state toggles $\ket{+} \leftrightarrow \ket{-}$ when it is phase-gated with data qubit in state $\ket{1}$, but it is unchanged if that qubit is $\ket{0}$; thus the final state is $\ket{-}$ if and only if there have been an odd number of such toggles.) Now in the present physical system, we can perform an operation that is essentially identical to the desired phase gate by exploiting the dipole-dipole interaction between the probe and the nearby data qubit. {In our scheme the separation between data qubits is at least 10 times greater than the probe-data separation, thus the interaction of the probe and the three data qubits to which it is not immediately proximal is three orders of magnitude weaker and can be treated as negligible, to an excellent approximation. Therefore the Hamiltonian of interest that of two $S=\frac{1}{2}$ spins, each in a static B field in the Z direction and experiencing a dipole-dipole interaction with one another, which is}
\[
H_{\rm 2S}=\mu_{B} B(g_{1}\sigma_1^z+g_{2}\sigma_2^z)+\frac{J}{r^3}\big( \boldsymbol{\sigma}_1\cdot \boldsymbol{\sigma}_2 - 3({\boldsymbol{\hat r}}\cdot \boldsymbol{\sigma}_1 )({\boldsymbol{\hat r}}\cdot \boldsymbol{\sigma}_2 )\big).
\]

Here $\boldsymbol{r}$ is the vector between the two spins, and $\boldsymbol{\hat r}$ is the unit vector in this direction and $J = \frac{\mu_0 g^2_e\mu^2_B}{4\pi}$. In the present analysis we assume that the Zeeman energy of the probe spin {\em differs} from the Zeeman energy of the data qubit by an amount $\Delta=\mu_{B}B(g_{1}-g_{2})$ which is orders of magnitude greater than the dipolar interaction strength $J/r^3$, a condition that prevents the spins from `flip-flopping', as shown in~\cite{Benjamin2004}. Then in a reference frame that subsumes the continuous Zeeman evolution of the spins their interaction is simply of the form 
\begin{equation}
\label{eqn:S}
\begin{aligned}
S(\theta)&=\left(\begin{array}{cccc}1 & 0 & 0 & 0 \\0 & \exp(i\theta) & 0 & 0 \\0 & 0 &  \exp(i\theta) & 0 \\0 & 0 & 0 & 1\end{array}\right) \\
&=\left(\cos\frac{\theta}{2}\right)I-i\left(\sin\frac{\theta}{2}\right) Z_1Z_2. 
\end{aligned}
\end{equation}
where the expressions discard irrelevant global phases, $I$ is the identity and $Z_1$, $Z_2$ are Pauli matrices action on the two spins respectively. 

{The condition that $\Delta\gg J/r^3$  will certainly be met if, as we suggest, the probe and data qubits are of different species. Suppose that the data qubits are phosphorus donors in silicon, while the probes are NV centres in diamond. The zero-field splitting of an NV centre is of order \SI{3}{\GHz}, while there is no equivalent splitting for the phosphorus donor qubit; this discrepancy implies $\Delta$ is more than six orders of magnitude greater than $J/r^3$ (the latter being of order \SI{0.8}{\kHz} at $r=40$~nm). A similar conclusion can be reached even if the probe and data qubits are both silicon-based, for example if each probe is a bismuth donor and each data qubit is a phosphorus donor. Given the hyperfine interactions strengths for phosphorus and bismuth donors of \SI{118}{\MHz} and \SI{1475}{\MHz} respectively, the typical minimum detuning between the two species is nearly six orders of magnitude greater than $J/r^3$. We performed exact numerical simulation of the spin-spin dynamics using these values, finding as expected that deviations from the form of $S(\theta)$ given above are extremely small, of order $10^{-4}$ or lower. }

For our purposes $U=S(\pi/2)$ is an ideal interaction: it is equivalent to the canonical two-qubit phase gate $G$ (up to an irrelevant global phase) if one additionally applies local single-qubit gates $V^{\dagger}=diag\{1,-i\}$ to each qubit. Thus the desired four-qubit parity measurement is achieved when the probe experiences an $S(\pi/2)$ with each qubit in turn, followed by measurement of the probe and application of $V^{\dagger}$ to all four data qubits (see Fig~\ref{interactionFig}(d)).

Our goal is therefore to acquire this maximum entangling value of $\theta=\pi/2$ during the time that the two spins interact. For the present paper we consider two basic possibilities for the way in which the mobile probe spin moves past each static  data qubit; these two cases are shown in Fig.~\ref{interactionFig}(b). The first possibility is that the probe moves abruptly from site to site, remaining stationary in close proximity to each data qubit in turn. In this case, we simply have $\theta=\alpha t$ where $\alpha = \frac{\mu_0 g^2_e\mu^2_B}{4\pi d^3}$ i.e. the phase acquired increases linearly until the probe jumps away. The motion of the probe between sites is assumed to be on a timescale that is very short compared to the dwell time at each site; in practice this motion might be in-plane or it might involve lifting and dropping the probe. 

\begin{figure*}[t]
\centering
\includegraphics[width=2.\columnwidth]{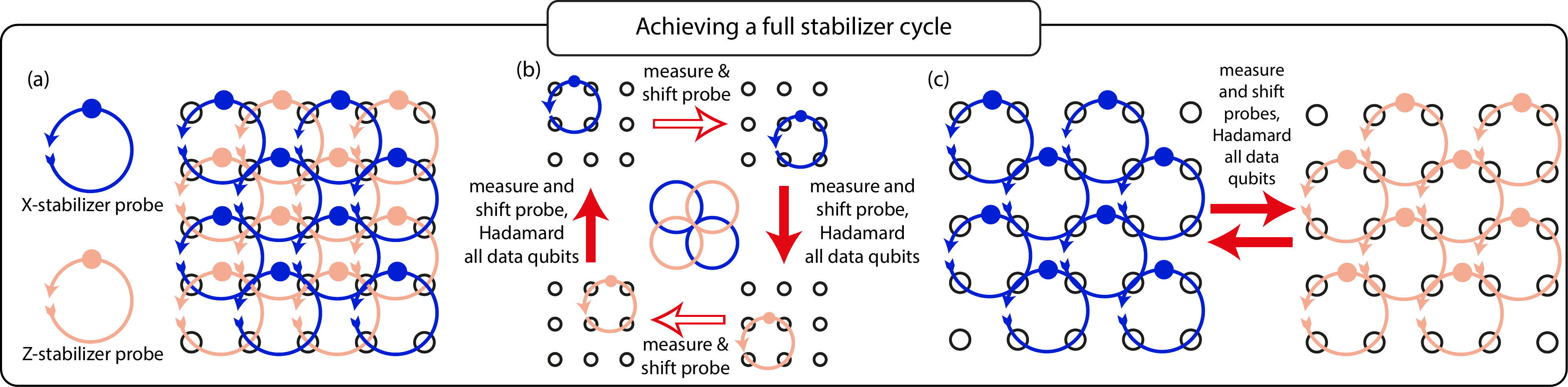}
\caption{\label{ProbeLayouts}\color{black}{Three approaches to implementing the full surface code. (a) The probe stage is manufactured with an identical lattice to the data qubits. In this approach all the X-stabilizer operations are performed in parallel, with the probes for the Z-stabilizers made ``inactive'' by preparation in the $\ket{0}$ state. A global Hadamard is then performed on the data qubits. Finally, the Z-stabilizers are all measured with the X-probes made inactive. The correction of the extra phase acquired by interaction with inactive probes can be subsumed into the global Hadamard operations. If the time for a probe to complete one orbit is $\tau$ then this approach takes $2\tau$ to complete a full round of stabilizer measurements. (b) The probe stage has 1/4 the qubit density of the data lattice. All probes are ``active'' (prepared in $\ket{+}$) throughout. A more complex probe orbit is required to achieve this approach: here a ``four-leaved clover'' motion. This protocol has time cost $\sim 4\tau$ per round. This is the approach simulated to produce our threshold results. (c) The probe stage is manufactured with 1/2 the qubit density of the data stage. An abrupt shift of the probe stage is required between the rounds of X- and Z-stabilizers. All probes are ``active'' throughout and this method requires time $\sim 2\tau$ per round.}}
\end{figure*}

An alternative which might be easier to realise is that the probe moves continuously with a circular motion (since this corresponds to in-phase simple harmonic motion of the probe stage in the $x$ and the $y$ directions). Because the data qubits are widely spaced, from the point of view of a data qubit the probe will come in from a great distance, pass close by and then retreat to a great distance. The interaction strength then varies with time; but by choosing the speed of the probe we can select the desired total phase shift, i.e. we again achieve $S(\pi/2)$. \textcolor{black}{The nature of the circular orbit has positive consequences in terms of tolerating implantation errors, as we presently discuss (see Fig.~\ref{resultsFig} upper panels versus lower panels). However, our simulations indicate the continuous circular motion does slow the operation of the device by approximately a factor of ten as compared to abrupt motion; therefore there is more time for unwanted in-plane spin-spin interactions to occur (see Appendix~\ref{Appendix::dipole background}). To compensate we may adjust the dimensions of the device, for example choosing $d=33$~nm with $D=700$~nm will negate the increase.}

\textcolor{black}{The analysis in this paper will establish that it suffices for our device to have local control only of the probe qubits, in order to prepare and to measure in the $X$, $Y$ or $Z$ basis \footnote{To achieve a probe state initialisation in the X or Y basis, we rely on initialisation in the Z eigenstate basis and subsequent spin rotations using the local probe control.}. Global control pulses suffice to manipulate the probe qubits during their cycles, and moreover the data qubits can be controlled entirely through global pulses. With these building blocks we can meet the surface code requirements of measuring four-qubit parity in both the $Z$ and the $X$ basis.} The surface code approach to fault tolerance requires one to measure parity in both the $Z$ and the $X$ basis. Crucially, both types of measurement can be achieved with the same probe cycle. The $X$-basis measurements differ from the $Z$-basis simply through the application of global Hadamard rotations to the data qubits before, and after, the probe cycle (see Fig.~\ref{ProbeLayouts} rightmost). Note that the additional phases $V^\dagger$ required in our protocol are easily accounted for in the scheme e.g. by adjusting the next series of Hadamard rotations to absorb the phase.

However the surface code protocol does require that the data qubits involved in X- and Z-basis stabilizers are grouped differently. This can be achieved in a number of ways; generally there is a tradeoff between the number of probe qubits required,  the time taken to complete one full round of stabilizer measurements, and the complexity of the motion of the moving stage. Three possible approaches are detailed in Fig.~\ref{ProbeLayouts}. The fastest protocol involves manufacturing identical probe and data grids. This approach also has the simplest implementation with regards to the mechanical motion of the moving stage: it can be performed with continuous circular movement. In this approach there must be a method of `deactivating' probes which are not presently involved in the parity measurements. This can simply be achieved by preparing these probes in the $\ket{0}$ states so that they do not entangle with the data qubits, instead only imparting an unconditional phase shift (the correction of which can be subsumed into the next global Hadamard cycle).

\textcolor{black}{The same `deactivation' of probes also allows us the flexibility to perform either three- or two-qubit stabilizers should we wish to, or indeed one-qubit stabilisers i.e. measurement of a specific data qubit. This can be achieved without altering the regular mechanical motion by appropriately timing the preparation and measurement of a given probe during its cycle: it should be in the deactivated state while passing any qubits that are not to be part of the stabiliser, but prepared in the $\ket{+}$ state prior to interacting with the first data qubit of interest. Probe measurement to determine the required stabiliser value should be performed in the $Y$-basis after interacting with one or with three data qubits, or in the $X$-basis after interacting with two data qubits. Note that the simple phase shifts induced by the `deactived' probe can either be tracked in the classical control software, or negated at the hardware level by repeating a cycle twice: once using $\ket{0}$ for the deactived probe and once using state $\ket{1}$.}

\textcolor{black}{Our simulations assume solution (b) from Fig.~\ref{ProbeLayouts}; this solution divides a full cycle into four stages, but has the considerable merit that it requires the fewest probes and therefore the lowest density for the measurement/initialisation systems.}


The description above is in terms of ideal behaviour, but we should consider a wide variety of defects and errors in order to establish whether the device is realistic with present or near-future technology. This includes not only the imperfections in state preparation, measurement and manipulation, but moreover also the systematic errors that result from the spins occupying positions that deviate from their ideal location. Given a full model of these errors, we can determine how severe the defects can be before the device ceases to operate as a fault tolerant quantum memory: this is the fault tolerance {\em threshold}. Presently we describe our numerical simulations which have determined this threshold.

In obtaining these results, we had to tackle a number of unusual features of this novel mechanical device. The most important point is that we must suppress the systematic errors that arise from fixed imperfections in the locations of the spins. Each data qubit is permanently displaced from its ideal location by a certain distance in some specific direction, and these details may be unknown to us -- what is the effect of such imperfections on the idealised process of four-qubit parity measurement described above? Our analytic treatment (see Appendix~\ref{Appendix:AnalyticSO}) reveals that the general result is to weight certain elements of the parity projection irregularly. Specifically, whereas the ideal even parity projector is  
\[{\hat P}_{\rm even}=\ket{0000}\bra{0000}+\ket{1100}\bra{1100}+...+\ket{1111}\bra{1111}
\]
when the spins involved are misaligned then one finds that different terms in the projector acquire different weights, so that the projector has the form,
\begin{eqnarray}
{\hat P}^\prime_{\rm even}=A\big(\ket{0000}\bra{0000}&+&\ket{1111}\bra{1111}\big)+\nonumber \\
B\big(\ket{0011}\bra{0011}&+&\ket{1100}\bra{1100}\big)+\nonumber \\
C\big(\ket{0110}\bra{0110}&+&\ket{1001}\bra{1001}\big)+\nonumber \\
D\big(\ket{0101}\bra{0101}&+&\ket{1010}\bra{1010}\big) + \hat W \nonumber 
\end{eqnarray}
where $\hat W$ is a set of lower weighted projectors on odd states. Meanwhile, the odd parity projector, ${\hat P}_{\rm odd}$ becomes ${\hat P}^\prime_{\rm odd}$ which is similarly formed of a sum of pairs; each pair such as $(\ket{0001}\bra{0001}+\ket{1110}\bra{1110})$ has its own weighting, differing from that of the other permutations. Using these projectors to measure the stabilizers of the surface code presents the problem that the error is systematic: for a particular set of four spins, the constants $A$, $B$, $C$ and $D$ will be the same each time we measure an `even' outcome. Each successive parity projection would enhance the asymmetry. In order to combat this effect, and effectively `smooth out' the irregularities in the superoperator, we introduce a simple protocol that is analogous to the `twirling' technique used in the literature on entanglement purification. Essentially we deliberately introduce some classical uncertainty into the process, as we now explain.

\begin{figure*}[t]
\centering
\includegraphics[width=2.\columnwidth]{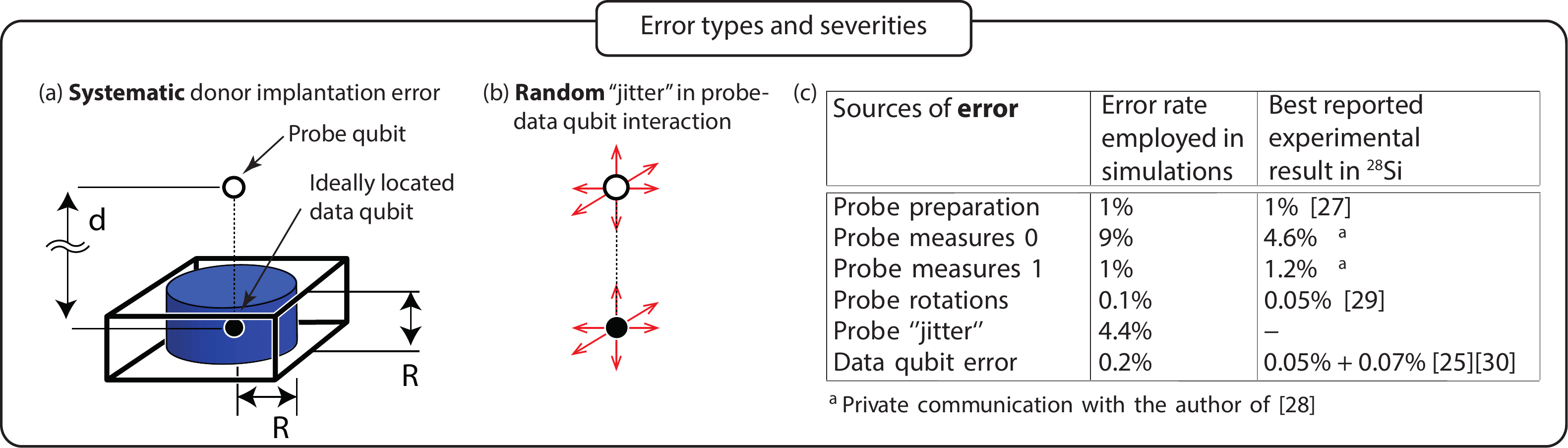}
\caption{\label{Sources of error}Our error model. (a) The imprecision in donor implantation within the silicon substrate is a systematic error; it is the same on each pass of the probe. We model this as each data qubit being randomly misplaced according to some distribution which will depend on the method used to manufacture the qubit arrays. Pictured here, a data qubit at a random fixed position with uniform probability inside the blue pillbox. (b) We also include a random fluctuation in the field strength of the dipole-dipole coupling, which we call ``jitter''. This would correspond to random spatial vibration of the probe qubit, or a random error in the timing of the orbit. This error occurs at each probe-data interaction independently. (c) Full table of the additional sources of error that are considered in our simulations and the experimental state-of-the-art for each in doped $^{28}\text{Si}$. In each case note that we select fidelities for our simulations which are comparable with those which have been experimentally demonstrated. See Appendix~\ref{Appendix:ErrorModel} for more details of the error model. \textcolor{black}{ Note the experimental numbers for data qubit error are computed from reported rates applied over $1.2\,\mathrm{ms}$, the cycle time given an abruptly moving probe and a $40\,\mathrm{nm}$ separation; achieving the same rates for the circular orbit would require improved materials and/or a smaller separation.  }}
\label{fig:errorTable}
\end{figure*}

Suppose that one were to apply the imperfect ${\hat P}^\prime_{\rm even}$ projector to four data qubits, but immediately prior to the projection and immediately after it we flip two of the qubits. For example, we apply $XX\Identity\Identity$ before and after, where $X$ is the Pauli $x$ operator and $\Identity$ is the identity. The net effect would still be to introduce (unwanted) weightings corresponding to $A$, $B$, $C$ and $D$, however these weights would be associated with different terms than for ${\hat P}^\prime_{\rm even}$ alone; for example the $A$ weight will be associated with $\ket{1100}$ and $\ket{0011}$. Therefore, consider the following generalisation: we randomly select a set of unitary single qubit flips to apply both before and after the ${\hat P}^\prime_{\rm even}$ projector, from a list of four choices such as  $U_1=\Identity\Identity\Identity\Identity$, $U_2=\Identity\Identity XX$, $U_3=\Identity X\Identity X$, $U_4=\Identity XX\Identity$. That is, we choose to perform our parity projection as $U_i {\hat P}^\prime_{\rm even} U_i$ where $i$ is chosen at random. We then note the parity outcome, `odd' or `even', and {\em forget} the $i$. The operators representing the net effect of this protocol, ${\hat P}^{\rm smooth}_{\rm even}$ and  ${\hat P}^{\rm smooth}_{\rm odd}$  are specified in Appendix~\ref{Appendix:AnalyticSO}. Essentially we replace the weightings $A$, $B$, $C$ and $D$ with a common weight that is their average, but at the cost of introducing  Pauli errors as well as retaining the problem that ${\hat P}^{\rm smooth}_{\rm even}$  has a finite probability of projecting onto the odd subspace. Analogously ${\hat P}^{\rm smooth}_{\rm odd}$ involves smoothed out odd projectors, newly introduced Pauli error terms, and a retained risk of projection onto the even subspace. However these imperfections are tolerable -- indeed they will occur in any case once we allow for the possibility of imperfect preparation, rotation, and measurement. Crucially, the `twirling' protocol allows us to describe the process in terms of a superoperator that we can classically simulate. It is formed from a probabilistically weighted sum of simple operators, each of which is either ${\hat P}_{\rm even}$  or ${\hat P}_{\rm odd}$, together with some set of single qubit Pauli operations, i.e. $S_1S_2S_3S_4$ where $S_i$ belongs to the set $\{\Identity,X,Y,Z\}$. In our simulation we can keep track of the state of the many-qubit system by describing it as the initial state together with the accumulated Pauli errors.

In practice it may be preferable to achieve an equivalent effect to the $U_i$ twirling operators {\em without} actually applying operations to the data qubits. This is possible since flipping the probe spin before-and-after it passes over a given data qubit is equivalent to flipping that data qubit, i.e. it is only the question of whether there is a net flip between the probe and the data qubit which affects the acquired phase. Therefore we can replace the protocol above with one in which the probe is subjected to a series of flips as it circumnavigates its four data qubits, while those data qubits themselves are not subjected to any flips. Since we are free to choose the same $i=1\ldots 4$ for all parity measurements occurring at a given time, these probe-flipping operations can be global over the device. In this approach the only operations that target the data qubits are the Hadamard rotations at the end of each complete parity measurement. This is an appealing picture given that we wish to minimise noise on data qubits, and it is this variant of the protocol which we use in our numerical threshold-finding simulations, the results of which are shown in Fig.~\ref{resultsFig}.

It is worth noting that in many real systems we may wish to use a spin echo technique to prevent the probe and data qubits from interacting with environmental spins. \textcolor{black}{In this case we would apply at least one flip to the spins (both the data and probe families simultaneously) during a parity measurement cycle; fortunately it is very natural to combine such echo flips with the flips required for twirling. The time for the parity measurement is thus not limited by the dephasing time $T_2^*$, but by the more generous coherence time $T_2$} \footnote{{\color{black} Using similar techniques, it is also possible to decouple certain parts of the dipole-dipole interaction between the probe and the data qubits, which would  reduce phase gate errors from probe qubits that are too strongly coupled to their data qubits. We note that the results of this paper do not make use of this performance improving technique.}}.

Before concluding our description of this protocol we note that the idea of perturbing our system and then deliberately {\em forgetting} which perturbation we have applied, whilst perfectly possible, will not be the best possible strategy. We speculate that superior performance would result from cycling systematically though the $U_i$ choosing $i=1,2,3,4,1,2...$ over successive rounds; but for the present paper is suffices to show that even our simple random twirl leads to fault tolerance with a good threshold.

\begin{figure*}[t]
\centering
\includegraphics[width=2.\columnwidth]{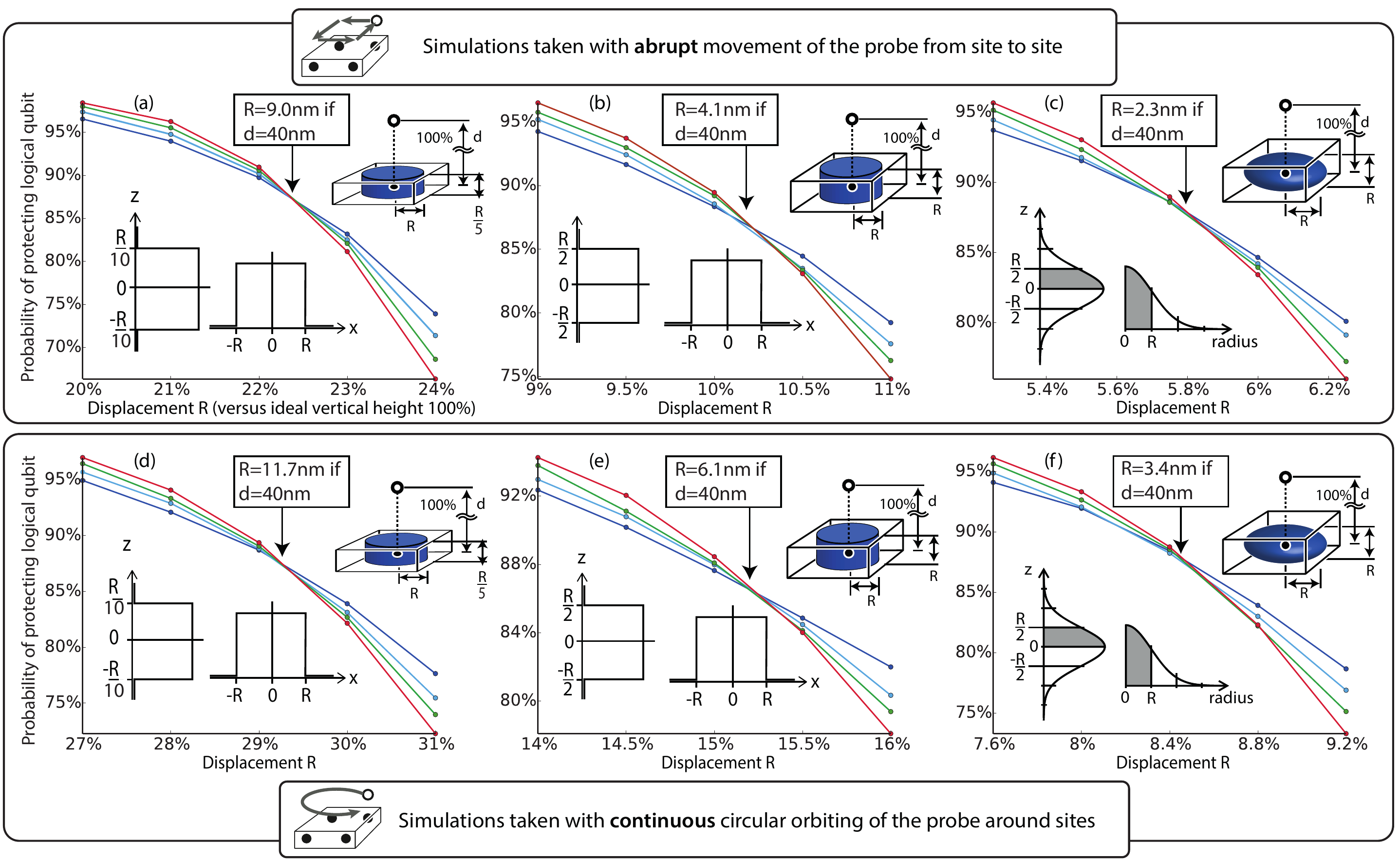}
\caption{\label{resultsFig}Results of threshold-finding numerical simulations. A system has surpassed the threshold for fault tolerant representation of a logical qubit if, when the system size is increased, we increase the probability of storing that logical qubit without corruption. Thus the crossing point of the lines reveals the threshold point. Lines from blue to red correspond to increasing array sizes of $221$, $313$, $421$ and $545$ physical qubits. Figure~\ref{Sources of error}(c) specifies the assumed error levels in preparation, control and measurement of the probe qubits. The data in (a) \& (d) are for the disk-shaped distribution shown in the inset -- data qubits are located with a circle of radius $R$ in the $x$-$y$ plane, and with a $z$ displacement $\pm R/10$. The data in (b) \& (e) are for a pillbox distribution, with a ratio of $2:1$ between lateral and vertical displacement. In (c) \& (f) the same $2:1$ ratio is used, but with a normal distribution where $R$ is now the standard deviation. Each data point in the figures corresponds to at least $50,000$ numerical experiments. Decoding is performed using the Blossom V implementation of Edmond's minimum weight perfect matching algorithm~\cite{Kolmogorov2009,Edmonds1965}.}
\end{figure*}

\bigskip
\bigskip
\noindent{\bf \large Results of numerical simulations}

We combined the protocol described above, whose purpose is to smooth out systematic errors, with an error model that accounts for finite rates of random error in the preparation, control, and measurement of the spins involved. The error model we employed is the standard one in which, with some probability $p$, an ideal operation is followed by an error event: a randomly selected Pauli error, or simple inversion of the recorded outcome in the case of measurement. We specify the model more precisely in Appendix~\ref{Appendix:ErrorModel}. {\color{black} In Fig.~\ref{Sources of error}(c) we tabulate the particular error probabilities used in our simulations, which are compatible with the values found in the literature for phosphorus in \textsuperscript{28}Si. \textcolor{black}{These values and possible methods of implementing these operations are discussed at length in the section ``Practicality of the device''.} The relative importance of the different errors is explored in additional simulations presented in Appendix~\ref{Appendix:furtherthresholds}.}

\textcolor{black}{In addition to these sources of error we investigate three particular models of the systematic misalignment of the data qubits and two different forms of probe orbit. The uniform disk-shaped distribution of donors, where error in the z-position of the donor is five times smaller than the lateral positional error, might be the error probability distribution one expects of a more sophisticated fabrication technique involving precise placement of donor in a surface via an STM tip with layers of silicon then grown over. The pillbox and normal distributions might be more representative of the results of using an ion implantation technique with very low or very high levels of straggle respectively. More discussion of the merits of these techniques can be found below under ``Practicality of the device''. The two probe orbits investigated can be considered two limiting cases: a circular orbit will be slower, with long transit times as the probe moves between data qubits in which nothing useful is happening, but this can be implemented with simple harmonic motion of the probe (or data) stage; the abrupt orbit, in which we imagine the probe jumps instantly from site to site, is faster but likely more difficult to achieve with micro-electromechanical systems. }

\textcolor{black}{Our threshold-finding simulation generates a virtual device complete with a specific set of misalignments in the spin locations (according to one of the distributions), a specific probe orbit and the additional sources of error details in  Fig.~\ref{Sources of error}(c).} It tests this device to see whether it successfully protects a logical qubit for a given period, and then repeats this process over a large number of virtual devices generated with the same average severity of misalignments. Thus the simulation determines the probability that the logical qubit is indeed protected in these circumstances. By performing such an analysis for devices of different size (i.e. different numbers of data qubits) we determine whether this particular set of noise parameters is within the threshold for fault tolerance -- if so, then larger devices will have superior noise suppression. Repeating this entire analysis for different noise parameters allows us to determine the threshold precisely. The results are shown in Fig.~\ref{resultsFig}, and are derived from over six million individual numerical experiments.

\textcolor{black}{The threshold results are shown in terms of qubit misplacement error as a percentage of the ideal probe-data separation $d$. The use of the long range dipole-dipole interaction means one can choose the scale of the device. On each plot we indicate the error tolerated for the specific case of $d=\SI{40}{nm}$.} These show an extremely generous threshold in the deviation in the positioning of the implanted qubits, with displacements of up to \SI{11.7}{nm} being tolerable in the best case scenario Fig.~\ref{resultsFig}(d). \textcolor{black}{Thus if donor qubits can be implanted with better accuracy than these values over a whole device, which otherwise operates with the errors detailed in Figure~\ref{Sources of error}(c), we can arbitrarily suppress the logical qubit error by increasing the size of the qubit grid.} We note that the continuous motion mode leads to a higher tolerance than the abrupt motion mode, as the smooth trajectory means a lower sensitivity to positional deviations in the $x$-$y$ plane.

\begin{figure}[t]
\centering
\includegraphics[width=\columnwidth]{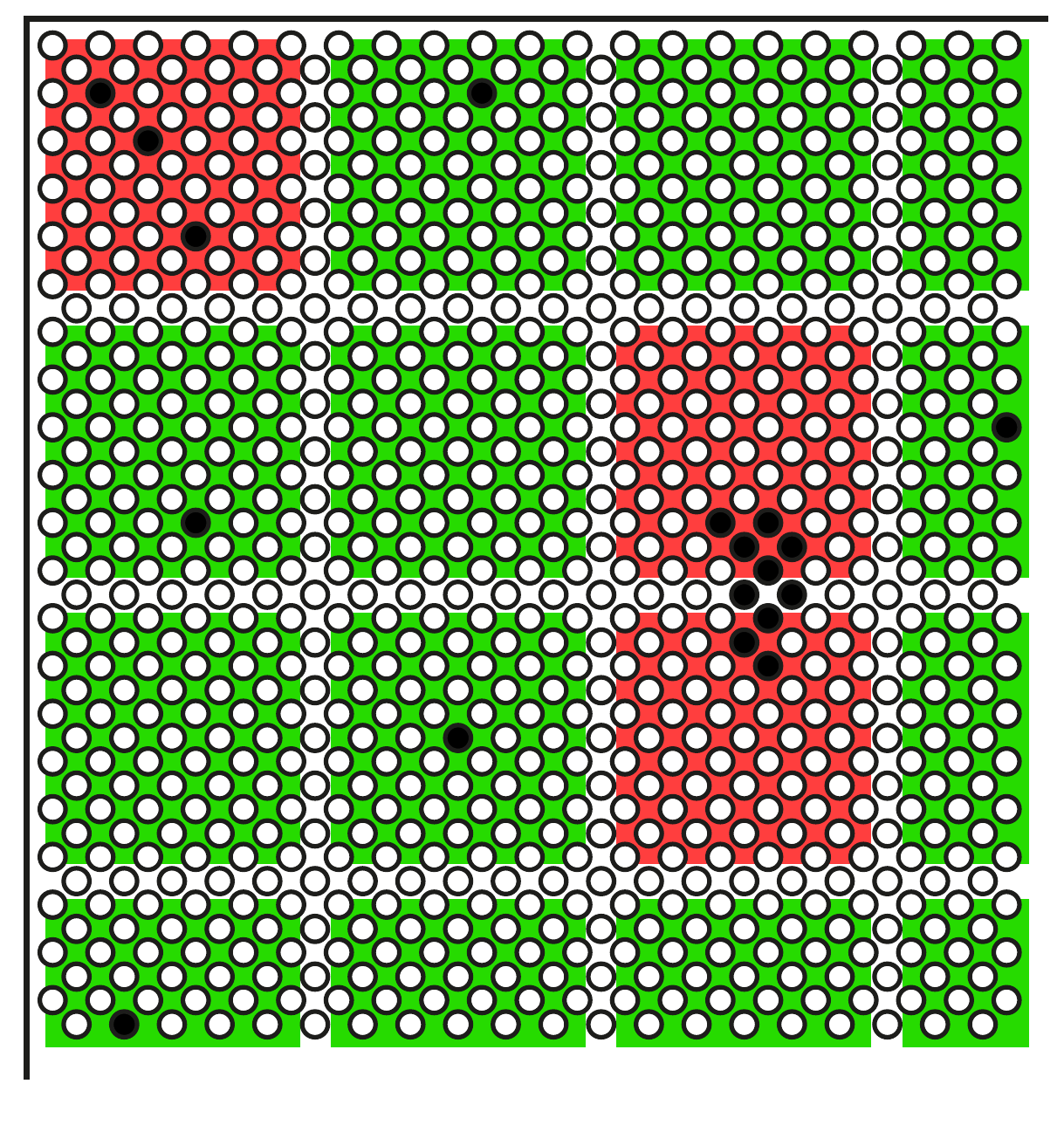}
\caption{Example of how one could encode and process multiple logical qubits into a flawed array. Each white circle is a data qubit; each green patch is a subarray representing a single logical qubit. When we wish to perform a gate operation between two logical qubits then we begin making parity measurements at their mutual boundary, according to the lattice surgery approach~\cite{latSurgery}. If a region of the overall array contains multiple damaged or missing data qubits, we simply opt not to use it (red patches). Note that in a real device the patch structure would probably be several times larger in order to achieve high levels of error suppression.
\label{multiQubit}}
\label{interleave}
\end{figure}

\bigskip
\noindent{\bf \large Generalisation to quantum computation}


Our simulations have found the threshold for a quantum {\em memory} using the surface code. It is conventional to take this as an accurate estimate of the threshold for full quantum computing using the surface code~\cite{Raussendorf2007}. The justification for this is non-obvious and we summarise it here. \textcolor{black}{Universal quantum computation requires a universal gate set. One such set is the Clifford group, augmented by a single non-Clifford operation - often the `$T$ gate' (the $\pi/8$ rotation) is chosen. Computation then involves, at the logical level, only the Clifford operations and this $T$ gate. A subset of the Clifford group can be performed fault-tolerantly within  the surface code, while logical X and Z operations can actually be implemented in software by updating the Pauli frame. Logical measurement of $\{\ket{0},\ket{1}\}$ can be achieved in a transversal way by simply measuring all the data qubits individually in the block, and the CNOT is also transversal in the surface code. The Hadamard gate is \textit{almost} transversal - when performed transversally the logical qubit is rotated, but this rotation can be `fixed' by a procedure of enforcing a slightly different set of stabilizers at the boundaries~\cite{latSurgery}, which we would achieve through `deactivating' probes as required. To complete the Clifford group the so-called `$S$ gate' ($\pi/4$ rotation) is required. Neither this nor the non-Clifford $T$ gate are directly supported in a surface code-based device; they can  be achieved by consuming an additional encoded qubit in a {\em magic state}~\cite{Bravyi2005}.}  

The fundamentally new element required for computing is \textcolor{black}{therefore} the creation of magic states. For our purposes there are two issues to confirm: Can we make such states within our architecture, and, does the need to do so revise our threshold? Magic state generation involves injection (mapping a single physical qubit to an encoded qubit) and distillation (improving the fidelity of such encoded states by sacrificially measuring some out). The latter involves only Clifford operations \textcolor{black}{(which may include a previously distilled $S$ gate)}, and therefore falls under the discussion in our previous paragraph. Injection requires operations on individual data qubits rather than the groups of four, but this is possible within our constraint of sending only global pulses to our data qubits: control of individual probe spins implies the ability to control individual data qubits and indeed to inject a magic state~\footnote{\color{black}{For example, suppose we set probes to $\ket{1}$ where we wish to have a net effect on the adjacent data qubits, and $\ket{0}$ elsewhere. Consider the sequence $Y(-\pi/2)Z(-\pi/8)S(\pi/8)Y(\pi/2)$, where $Y$ and $Z$ are Pauli rotations both performed on all data qubits and $S$ is the probe-data qubit interaction, Eqn.~\ref{eqn:S}. The net effect on a data qubit is the identity when the probe is in $\ket{0}$, but when the probe is $\ket{1}$ it constitutes a rotation of $X(\pi/4)$ on the data qubit, where is $X$ is the Pauli rotation. This operation would take a data qubit from $\ket{0}$ to a magic state $\cos(\pi/8)\ket{0}-i\sin(\pi/8)\ket{1}$ if-and-only-if the proximal probe is in state $\ket{1}$.}}. Crucially, the purification threshold for a noisy magic state is known to be much larger than that for a surface code memory; indeed this threshold has just been further relaxed in a recent study~\cite{Li2015}. Ultimately, therefore, the overall threshold for quantum computation is indeed set by the memory threshold, as discussed by the relatively early literature~\cite{Raussendorf2007}.

More generally, one can ask about how multiple qubits should be encoded into a large array of data qubits, and what the impact of flaws such as missing data qubits would be on a computation. While a detailed analysis lies beyond the scope of the present paper, approaches such as lattice surgery~\cite{latSurgery} can offer one simple solution that is manifestly tolerant of a finite density of flaws. The approach is illustrated in Fig.~\ref{multiQubit}. Square patches of the overall array are assigned to hold specific logical qubits; stabilizers are not enforced (i.e. parity measurements are not made) along the boundaries except when we wish to perform an operation between adjacent logical qubits. Importantly, if a given patch is seriously flawed (because of multiple missing data qubits during device synthesis, or for other reasons) then we can simply opt not to use it -- it becomes analogous to a `dead pixel' in a screen or CCD. As long as such dead pixels are sufficiently sparse, then we will always be able to route information flow around them.

\bigskip
{\bf \noindent \large Practicality of the device}
\smallskip

We now discuss the practicality of the proposed device, in light of the robustness to defects that our simulations have established.

\subsection*{Timescales and decoherence}

First, we examine the operational timescale of the device: For the aforementioned probe - data qubit separation of $d = 40\,\mathrm{nm}$, the total interaction time for the four $S(\pi/2)$ phase gates with the four data qubits is  $t_\mathrm{int} = 2\pi d^3/J \approx 1.2\,\mathrm{ms}$. {\color{black} This stabilizer cycle time was chosen because it is short enough to comply with the coherence times of donors in silicon or the NV center and long enough to avoid a significant operational lag due to the finite operation speed of the stage (see below)}. In the abrupt motion scenario Fig. \ref{interactionFig}(b)(i) with negligible transfer times, this would allow an operation of the device at about \SI{1}{\kHz}. In the continuous circular motion picture, a significant time is required for the probe's transfer between data qubits which slows down the device by roughly a factor of $D/d=10$. \textcolor{black}{As noted earlier and discussed in more detail in Appendix~\ref{Appendix::dipole background}, the consequent increased accumulation of unwanted in-plane spin-spin interactions can be negated by varying the dimensions, for example to $d = 33\,\mathrm{nm}$, $D = 700\,\mathrm{nm}$.} It is also worthwhile to consider a hybrid mode of operation with slow continuous motion in the vicinity of the data qubits  and fast accelerated transfers, as an approach that could provide both fast operation and high positional error tolerance. We further note that -- due to the $1/d^3$ dependence of the dipolar interaction -- every reduction of $d$ by a factor of two allows eight times faster operation frequencies. So, if manufacturing precision continues to improve as seen in the past, this device should be readily scalable to faster operations. In practice, $D/d$ may take a range of values and optimising it will be a trade off between the smallest possible fabrication feature sizes, the achievable translation velocities and the decoherence time of the qubits.


\subsection*{Mechanics and device design} 

The prototypical mechanical system that enables mechanical motions with sub-nanometer positional accuracy is the tip of an atomic force microscope cantilever. In principle, an array of tips on a single cantilever could incorporate the probe qubits and a cyclic motion of the cantilever would allow the four qubit phase gates.  Practical constraints such as  height uniformity of the probe tips, however, impose severe challenges on the scalability of this approach up to larger qubit grid sizes.

A more viable mechanical system could be represented by $x$-$y$ translation stages realised by micro electromechanical systems (MEMS). These devices are often manufactured from silicon-on-insulator wafers and could exploit the uniformity of the oxide layer to achieve a high homogeneity of the probe -- data qubit separation $d$ across the grid. Various  designs for MEMS $x$-$y$ translation stages have been put forward with travel ranges in the $\si{10 \micro\metre}$ range or higher \cite{Harness2000, Mukhopadhyay2008, Dong2007} and positional accuracies in the nm regime \cite{Koo2012, Chu2003}, both meeting essential requirements of our proposal. The motion speed of these stages is limited by their eigenfrequency and designs with frequencies $> 10\,\mathrm{kHz}$ \cite{Liao2010} permit stabilizer cycle translation times on the order of $\sim \SI{100}{\micro\second}$. 

\textcolor{black}{
Since the probe qubits need to be individually controlled and measured, local electric gates are required. There are two basic strategies depending on whether it is the probe array or the data qubit array that is in motion (recall that either can selected as the moving part; only the relative motion is significant). If the probe stage is mobile, then it is necessary to deliver the electrical contacts over the suspension beams at the side of the probe stage, by selecting beam characteristics appropriately~\footnote{To circumvent wide beams to incorporate all control leads, which would result in a large in-plane spring constant for the stage motion, we suggest using a multitude of thin beams. The spring constant of the beams for the in-plane motion of the stage scale with the width cubed. $N$ thin beams of width $w$ thus only increase the spring constant by a factor of $Nw^3$ compared to $(Nw)^3$ in the single wide beam case.}. If however  the data qubit grid forms the movable stage, then there is no need for such bridging since the data qubits are controlled purely through global pulses. The control for the probe qubits in the static silicon stage below could then be written in the same process as the probe spins themselves, relying on atomic-precise fabrication of phosphorous impurity SETs and gates (see below, \cite{Fuechsle2012, Hile2015}).}

\subsection*{Material systems}
We next direct our discussion to the properties of the proposed solid state qubit systems for this orbital probe architecture, namely donor impurities in silicon, the NV centers in diamond and divacancy centers in silicon carbide.

{\color{black} To achieve the aforementioned individual addressability of the probe qubits, either probe spins could be individually Stark shifted into resonance with a globally applied microwave source \cite{Pica2014, Laucht2015}, or, alternatively, magnetic field gradients could be applied to detune the individual resonance frequencies within the probe qubit grid.} Most of these systems also exhibit hyperfine structure, meaning that the transition energy between the $\ket{0}$ and $\ket{1}$ state of the electron spin depends on the nuclear spin state. This effectively results in two or more possible `species' of qubits, each of which must be manipulated by a microwave pulse of a different frequency. To perform the required qubit operations regardless of the nuclear spin state, we propose to use multi-tone microwave pulses composed of all resonance frequencies of the different species. With this we can ensure that nuclear spin flips - so long as they occur less frequently than the time for a stabilizer cycle - do not affect our ability to implement the proposed protocol. 

The following paragraphs discuss both the qubit performance of these systems and the suitability of the host material in MEMS applications.

\paragraph{\bf Donors in silicon}

Due to advanced fabrication processes and its excellent material properties, silicon is the predominant material for the realisation of high-quality MEMS devices. Furthermore, silicon  can be isotopically purified to a high degree, which reduces the concentration of \textsuperscript{29}Si nuclear spins and creates an almost ideal, spin-free host system. Consequently, electron spins of donor impurities, such as phosphorus, show extraordinarily long coherence times of up to \SI{2}{\s} \cite{Tyryshkin2012}, thus enabling a very low data qubit memory error probability (sub $0.1\%$) over the timescale of a single parity measurement of \SI{1.2}{\ms}. \textcolor{black}{If donors in silicon are employed as the probe qubits, then  initialisation and read-out of the electron spin of single dopants could be performed using spin-dependent tunnelling to a nearby reservoir and subsequent charge detection using SETs \cite{Morello2010, Pla2012, Muhonen2014}}. The average measurement fidelity is reported as \SI{97}{\percent} with read-out timescales on the order of milliseconds \textcolor{black}{\footnote{Although the device is tolerates measurement errors quite well, it should be noted that the read-out fidelity can be further increased if, by the end of a stabilizer cycle, the electron spin state is transferred to the nuclear spin \cite{Morton2008} from which it can then be measured with higher fidelity up to values of \SI{99.99}{\percent} \cite{Muhonen2014}. This advantage has to be traded against the cost of a longer measurement time: reported times of order \SI{100}{ms} are two orders of magnitude slower than other timescales described here, although faster readout of the electron spin with the help of optical-assisted ionisation of the donor may improve the timescale dramatically~\cite{Lo2015}}}. \textcolor{black}{The single qubit control fidelity for an electron spin of a single P donor in \textsuperscript{28}Si in these devices has been reported as \SI{99.95}{\percent}~\cite{Muhonen2015a}, which could be improved even further by the use of composite microwave pulses (as in \cite{Morton2005}).} Furthermore, it was shown in \cite{Muhonen2014} that the decoherence time of the qubit is not significantly affected by its proximity to the interface and can reach values up to \SI{0.56}{\s} with dynamical decoupling sequences. 

The footprint of the required electronic components to measure a single donor spin in silicon is typically on the order of $200 \times 200\,\mathrm{nm}^2$ and is thus small enough to achieve qubit grid separations of $D = 400\,\si{nm}$. \textcolor{black}{We note that if the required measurement temperatures on the order of \SI{100}{\milli\kelvin} become difficult to maintain, due to actuation motion and friction for example, then there are alternative optically assisted spin-to-charge conversion methods which may allow for single spin detection at liquid helium temperatures~\cite{Lo2015}.}

As shown by our threshold calculations, a key figure for this scheme is the implantation accuracy required for the probe and data qubit arrays. Ion implantation methods with resolutions approaching \SI{10}{\nm} can be achieved using either e-beam lithography directly on the substrate \cite{Vieu2000, Toyli2010} or nanostencil masks drilled into AFM cantilevers \cite{Weis2008}. For donors in silicon these approaches can be combined with ion impact detection to ensure deterministic single qubit implantation \cite{Jamieson2005}. Another technique for silicon is the STM tip patterning of a hydrogen mask and the subsequent exposure to phosphene gas, which enables atomically-precise ($\pm \SI{3.8}{\angstrom}$) phosphorus donor incorporation in all three dimensions \cite{Schofield2003, Fuechsle2012}. This accuracy is more than an order of magnitude below our calculated thresholds of Fig. \ref{resultsFig} and the challenge remaining is to maintain this precision over larger qubit arrays. 

\paragraph{\bf Diamond nitrogen-vacancy centers}
The electron spin qubit associated with the nitrogen-vacancy (NV) defect center of diamond features optically addressable spin states, which could be manipulated even at room temperature. \textcolor{black}{By using resonant laser excitation and detection of luminescence photons, fast (\SI{\sim 40}{\micro\s} \cite{Robledo2011}) and reliable (measurement fidelity of \SI{96.3}{\percent} \cite{Pfaff2014}) read-out of single NV center spins could be employed for the probe spin measurement and initialisation \footnote{ The nuclear spin of \textsuperscript{14}N or of adjacent \textsuperscript{13}C may again be exploited to enhance the measurement fidelity (\SI{99.6}{\percent} \cite{Waldherr2014}).}}. The coherence times in isotopically purified diamond samples ($T_2 = \SI{600}{ms}$ at \SI{77}{K} using strong dynamical decoupling \cite{BarGill2013}) are long enough to allow millisecond long stabilizer cycles \textcolor{black}{ and individual probe control using a global microwave field is similar to donors in silicon possible using electric gates and the Stark effect \cite{Dolde2011}}.

While the qubit operations possible in NV centers are advanced, so far very few micro-electromechanical devices have been realised using diamond. Among them are resonator structures from single crystalline diamond-on-insulator wafers  \cite{Ovartchaiyapong2012} and from nano-crystalline diamond \cite{Imboden2013}.  In principle though, diamond possesses promising material properties for MEMS applications \cite{Williams2011} and, given further research, could become an established material to build translatory stages.

The implantation accuracy for the NV center is determined by the ion beam techniques discussed above. The most accurate method uses a hole in an AFM cantilever and  achieves lateral accuracies of $\sim 25\,\mathrm{nm}$  at implantation depths of $8 \pm 3\,\mathrm{nm}$ \cite{Pezzagna2010}. This precision is only slightly below threshold of our scheme and it is reasonable to hope that new implantation methods could meet the requirements in the near future. Furthermore, we note that the proposed grid spacing of $D=400\,\mathrm{nm}$ is well beyond the diffraction limit for optical read-out ($250\,\mathrm{nm}$ \cite{Lesik2013}).

Another critical factor for all NV centre fabrication methods is the low yield of active NV centres per implanted nitrogen atom, which is typically well below \SI{30}{\percent} \cite{Naydenov2010}. \textcolor{black}{Such a low yield would result in too great a number of `dead pixels' within the layout specified in Fig.~\ref{multiQubit} to allow for the construction of a useful device.} 

While there are still significant challenges remaining to an integrated diamond MEMs probe array, it is encouraging that the basic requirement of our proposed scheme,  i.e. the control of the dipolar interaction of two electron spins by means of changing their separation mechanically, has already been achieved. Grinolds \textit{et al.} were able to sense the position and the dipolar field of a single NV centre by scanning a second NV centre in a diamond pillar attached to an AFM cantilever across it ---  at a NV centre separation of \SI{50}{\nm}  \cite{Grinolds2013}.

\paragraph{\bf Silicon carbide vacancy defects}
In addition to NV centres, divacancy defects of certain silicon carbide (SiC) polytypes exhibit optically addressable spin states suitable for qubit operations \cite{Koehl2011}. Furthermore, SiC micro electromechanical devices \cite{Adachi2013} and the required fabrication techniques have evolved in recent years, which could open up the possibility of a material with both optical qubit read-out and scalable fabrication techniques. Some important aspects of qubit operation, however, such as longer decoherence times (\SI{1.2}{\milli\second} reported in \cite{Christle2014}), single shot qubit read-out and deterministic defect creation with high positional accuracy have yet to be demonstrated.

\bigskip
\noindent{\bf \large Conclusion}

We have described a new scheme for implementing surface code quantum computing, based on an array of donor spins in silicon, which can be seen as a reworking of the Kane proposal to incorporate an inbuilt method for error correction. The required parity measurements can be achieved using continuous phase acquisition onto another `probe' qubit, removing the challenging requirement for direct gating between physical qubits. Through simulations using error rates for state preparation, control and measurement that are consistent with reported results in the literature, we find that this approach is extremely robust against deviations in the location of the qubits, with tolerances orders of magnitude greater than those seen in the origin Kane proposal. An additional benefit is that such a system is essentially scale independent, since the scheme is based on long range dipole interactions, so the dimensions of the device can be selected to match the available fabrication capabilities.

\bigskip

Acknowledgements.
Computing facilities were provided by the Oxford's ARC service, the Imperial College High Performance Computing Service, and the Dieter Jaksch group. The authors would like to acknowledge helpful conversations with Richard Syms. This work is supported by the EPSRC through a platform grant (EP/J015067/1) and the UNDEDD project (EP/K025945/1), and by the European Research Council under the European Community's Seventh Framework Programme (FP7/2007-2013) / ERC grant agreement no. 279781.  J.J.L.M. is supported by the Royal Society. The authors declare that the have no competing financial interests.

\bibliographystyle{apsrev4-1}
\bibliography{orbitalProbeBibNPJ}

\begin{thebibliography}{52}%
\makeatletter
\providecommand \@ifxundefined [1]{%
 \@ifx{#1\undefined}
}%
\providecommand \@ifnum [1]{%
 \ifnum #1\expandafter \@firstoftwo
 \else \expandafter \@secondoftwo
 \fi
}%
\providecommand \@ifx [1]{%
 \ifx #1\expandafter \@firstoftwo
 \else \expandafter \@secondoftwo
 \fi
}%
\providecommand \natexlab [1]{#1}%
\providecommand \enquote  [1]{``#1''}%
\providecommand \bibnamefont  [1]{#1}%
\providecommand \bibfnamefont [1]{#1}%
\providecommand \citenamefont [1]{#1}%
\providecommand \href@noop [0]{\@secondoftwo}%
\providecommand \href [0]{\begingroup \@sanitize@url \@href}%
\providecommand \@href[1]{\@@startlink{#1}\@@href}%
\providecommand \@@href[1]{\endgroup#1\@@endlink}%
\providecommand \@sanitize@url [0]{\catcode `\\12\catcode `\$12\catcode
  `\&12\catcode `\#12\catcode `\^12\catcode `\_12\catcode `\%12\relax}%
\providecommand \@@startlink[1]{}%
\providecommand \@@endlink[0]{}%
\providecommand \url  [0]{\begingroup\@sanitize@url \@url }%
\providecommand \@url [1]{\endgroup\@href {#1}{\urlprefix }}%
\providecommand \urlprefix  [0]{URL }%
\providecommand \Eprint [0]{\href }%
\providecommand \doibase [0]{http://dx.doi.org/}%
\providecommand \selectlanguage [0]{\@gobble}%
\providecommand \bibinfo  [0]{\@secondoftwo}%
\providecommand \bibfield  [0]{\@secondoftwo}%
\providecommand \translation [1]{[#1]}%
\providecommand \BibitemOpen [0]{}%
\providecommand \bibitemStop [0]{}%
\providecommand \bibitemNoStop [0]{.\EOS\space}%
\providecommand \EOS [0]{\spacefactor3000\relax}%
\providecommand \BibitemShut  [1]{\csname bibitem#1\endcsname}%
\let\auto@bib@innerbib\@empty
\bibitem [{our()}]{ourCode}%
  \BibitemOpen
  \href@noop {} {\ }\Eprint {http://arxiv.org/abs/Please see
  ``$/$naominickerson$/$fault$\_$tolerance$\_$simulations$/$releases'' on
  GitHub for the code we employed} {Please see
  ``$/$naominickerson$/$fault$\_$tolerance$\_$simulations$/$releases'' on
  GitHub for the code we employed} \BibitemShut {NoStop}%
\bibitem [{\citenamefont {Kane}(1998)}]{kane98}%
  \BibitemOpen
  \bibfield  {author} {\bibinfo {author} {\bibfnamefont {B.~E.}\ \bibnamefont
  {Kane}},\ }\href@noop {} {\bibfield  {journal} {\bibinfo  {journal} {Nature}\
  }\textbf {\bibinfo {volume} {393}},\ \bibinfo {pages} {133} (\bibinfo {year}
  {1998})}\BibitemShut {NoStop}%
\bibitem [{\citenamefont {Hollenberg}\ \emph {et~al.}(2006)\citenamefont
  {Hollenberg}, \citenamefont {Greentree}, \citenamefont {Fowler},\ and\
  \citenamefont {Wellard}}]{Hollenberg2006}%
  \BibitemOpen
  \bibfield  {author} {\bibinfo {author} {\bibfnamefont {L.~C.~L.}\
  \bibnamefont {Hollenberg}}, \bibinfo {author} {\bibfnamefont {A.~D.}\
  \bibnamefont {Greentree}}, \bibinfo {author} {\bibfnamefont {A.~G.}\
  \bibnamefont {Fowler}}, \ and\ \bibinfo {author} {\bibfnamefont {C.~J.}\
  \bibnamefont {Wellard}},\ }\href
  {http://link.aps.org/doi/10.1103/PhysRevB.74.045311} {\bibfield  {journal}
  {\bibinfo  {journal} {Physical Review B}\ }\textbf {\bibinfo {volume} {74}},\
  \bibinfo {pages} {045311} (\bibinfo {year} {2006})}\BibitemShut {NoStop}%
\bibitem [{\citenamefont {Schofield}\ \emph {et~al.}(2003)\citenamefont
  {Schofield}, \citenamefont {Curson}, \citenamefont {Simmons}, \citenamefont
  {Rue\ss{}}, \citenamefont {Hallam}, \citenamefont {Oberbeck} \emph
  {et~al.}}]{Schofield2003}%
  \BibitemOpen
  \bibfield  {author} {\bibinfo {author} {\bibfnamefont {S.~R.}\ \bibnamefont
  {Schofield}}, \bibinfo {author} {\bibfnamefont {N.~J.}\ \bibnamefont
  {Curson}}, \bibinfo {author} {\bibfnamefont {M.~Y.}\ \bibnamefont {Simmons}},
  \bibinfo {author} {\bibfnamefont {F.~J.}\ \bibnamefont {Rue\ss{}}}, \bibinfo
  {author} {\bibfnamefont {T.}~\bibnamefont {Hallam}}, \bibinfo {author}
  {\bibfnamefont {L.}~\bibnamefont {Oberbeck}},  \emph {et~al.},\ }\href
  {\doibase 10.1103/PhysRevLett.91.136104} {\bibfield  {journal} {\bibinfo
  {journal} {Phys. Rev. Lett.}\ }\textbf {\bibinfo {volume} {91}},\ \bibinfo
  {pages} {136104} (\bibinfo {year} {2003})}\BibitemShut {NoStop}%
\bibitem [{\citenamefont {Fuechsle}\ \emph {et~al.}(2012)\citenamefont
  {Fuechsle}, \citenamefont {Miwa}, \citenamefont {Mahapatra}, \citenamefont
  {Ryu}, \citenamefont {Lee}, \citenamefont {Warschkow} \emph
  {et~al.}}]{Fuechsle2012}%
  \BibitemOpen
  \bibfield  {author} {\bibinfo {author} {\bibfnamefont {M.}~\bibnamefont
  {Fuechsle}}, \bibinfo {author} {\bibfnamefont {J.~A.}\ \bibnamefont {Miwa}},
  \bibinfo {author} {\bibfnamefont {S.}~\bibnamefont {Mahapatra}}, \bibinfo
  {author} {\bibfnamefont {H.}~\bibnamefont {Ryu}}, \bibinfo {author}
  {\bibfnamefont {S.}~\bibnamefont {Lee}}, \bibinfo {author} {\bibfnamefont
  {O.}~\bibnamefont {Warschkow}},  \emph {et~al.},\ }\href
  {http://dx.doi.org/10.1038/nnano.2012.21} {\bibfield  {journal} {\bibinfo
  {journal} {Nat Nano}\ }\textbf {\bibinfo {volume} {7}},\ \bibinfo {pages}
  {242} (\bibinfo {year} {2012})}\BibitemShut {NoStop}%
\bibitem [{\citenamefont {E.~Dennis}\ and\ \citenamefont
  {Preskill}(2002)}]{topolo}%
  \BibitemOpen
  \bibfield  {author} {\bibinfo {author} {\bibfnamefont {A.~L.}\ \bibnamefont
  {E.~Dennis}, \bibfnamefont {A.~Kitaev}}\ and\ \bibinfo {author}
  {\bibfnamefont {J.}~\bibnamefont {Preskill}},\ }\href@noop {} {\bibfield
  {journal} {\bibinfo  {journal} {J. Math. Phys.}\ }\textbf {\bibinfo {volume}
  {43}},\ \bibinfo {pages} {4452} (\bibinfo {year} {2002})}\BibitemShut
  {NoStop}%
\bibitem [{\citenamefont {Fowler}\ \emph {et~al.}(2009)\citenamefont {Fowler},
  \citenamefont {Stephens},\ and\ \citenamefont
  {Groszkowski}}]{FOWLER_2009high}%
  \BibitemOpen
  \bibfield  {author} {\bibinfo {author} {\bibfnamefont {A.~G.}\ \bibnamefont
  {Fowler}}, \bibinfo {author} {\bibfnamefont {A.~M.}\ \bibnamefont
  {Stephens}}, \ and\ \bibinfo {author} {\bibfnamefont {P.}~\bibnamefont
  {Groszkowski}},\ }\href@noop {} {\bibfield  {journal} {\bibinfo  {journal}
  {Physical Review A}\ }\textbf {\bibinfo {volume} {80}},\ \bibinfo {pages}
  {052312} (\bibinfo {year} {2009})}\BibitemShut {NoStop}%
\bibitem [{\citenamefont {Wang}\ \emph {et~al.}(2011)\citenamefont {Wang},
  \citenamefont {Fowler},\ and\ \citenamefont
  {Hollenberg}}]{WANG_QCwithNNinteractionsAndErrorRatesOverOnePercent}%
  \BibitemOpen
  \bibfield  {author} {\bibinfo {author} {\bibfnamefont {D.~S.}\ \bibnamefont
  {Wang}}, \bibinfo {author} {\bibfnamefont {A.~G.}\ \bibnamefont {Fowler}}, \
  and\ \bibinfo {author} {\bibfnamefont {L.~C.~L.}\ \bibnamefont
  {Hollenberg}},\ }\href {http://dx.doi.org/10.1103/PhysRevA.83.020302}
  {\bibfield  {journal} {\bibinfo  {journal} {Phys. Rev.}\ }\textbf {\bibinfo
  {volume} {83}} (\bibinfo {year} {2011})}\BibitemShut {NoStop}%
\bibitem [{\citenamefont {Benjamin}\ and\ \citenamefont
  {Bose}(2004)}]{Benjamin2004}%
  \BibitemOpen
  \bibfield  {author} {\bibinfo {author} {\bibfnamefont {S.~C.}\ \bibnamefont
  {Benjamin}}\ and\ \bibinfo {author} {\bibfnamefont {S.}~\bibnamefont
  {Bose}},\ }\href {\doibase 10.1103/PhysRevA.70.032314} {\bibfield  {journal}
  {\bibinfo  {journal} {Phys. Rev. A}\ }\textbf {\bibinfo {volume} {70}},\
  \bibinfo {pages} {032314} (\bibinfo {year} {2004})}\BibitemShut {NoStop}%
\bibitem [{\citenamefont {Kolmogorov}(2009)}]{Kolmogorov2009}%
  \BibitemOpen
  \bibfield  {author} {\bibinfo {author} {\bibfnamefont {V.}~\bibnamefont
  {Kolmogorov}},\ }\href {http://dx.doi.org/10.1007/s12532-009-0002-8}
  {\bibfield  {journal} {\bibinfo  {journal} {Math. Prog. Comp.}\ }\textbf
  {\bibinfo {volume} {1}},\ \bibinfo {pages} {43} (\bibinfo {year}
  {2009})}\BibitemShut {NoStop}%
\bibitem [{\citenamefont {Edmonds}(1965)}]{Edmonds1965}%
  \BibitemOpen
  \bibfield  {author} {\bibinfo {author} {\bibfnamefont {J.}~\bibnamefont
  {Edmonds}},\ }\href@noop {} {\bibfield  {journal} {\bibinfo  {journal} {J.
  Math.}\ }\textbf {\bibinfo {volume} {17}},\ \bibinfo {pages} {449} (\bibinfo
  {year} {1965})}\BibitemShut {NoStop}%
\bibitem [{\citenamefont {Horsman}\ \emph {et~al.}(2012)\citenamefont
  {Horsman}, \citenamefont {Fowler}, \citenamefont {Devitt},\ and\
  \citenamefont {Meter}}]{latSurgery}%
  \BibitemOpen
  \bibfield  {author} {\bibinfo {author} {\bibfnamefont {C.}~\bibnamefont
  {Horsman}}, \bibinfo {author} {\bibfnamefont {A.~G.}\ \bibnamefont {Fowler}},
  \bibinfo {author} {\bibfnamefont {S.}~\bibnamefont {Devitt}}, \ and\ \bibinfo
  {author} {\bibfnamefont {R.~V.}\ \bibnamefont {Meter}},\ }\href
  {http://stacks.iop.org/1367-2630/14/i=12/a=123011} {\bibfield  {journal}
  {\bibinfo  {journal} {New Journal of Physics}\ }\textbf {\bibinfo {volume}
  {14}},\ \bibinfo {pages} {123011} (\bibinfo {year} {2012})}\BibitemShut
  {NoStop}%
\bibitem [{\citenamefont {Raussendorf}\ and\ \citenamefont
  {Harrington}(2007)}]{Raussendorf2007}%
  \BibitemOpen
  \bibfield  {author} {\bibinfo {author} {\bibfnamefont {R.}~\bibnamefont
  {Raussendorf}}\ and\ \bibinfo {author} {\bibfnamefont {J.}~\bibnamefont
  {Harrington}},\ }\href {http://dx.doi.org/10.1103/PhysRevLett.98.190504}
  {\bibfield  {journal} {\bibinfo  {journal} {Physical Review Letters}\
  }\textbf {\bibinfo {volume} {98}},\ \bibinfo {pages} {190504} (\bibinfo
  {year} {2007})}\BibitemShut {NoStop}%
\bibitem [{\citenamefont {Bravyi}\ and\ \citenamefont
  {Kitaev}(2005)}]{Bravyi2005}%
  \BibitemOpen
  \bibfield  {author} {\bibinfo {author} {\bibfnamefont {S.}~\bibnamefont
  {Bravyi}}\ and\ \bibinfo {author} {\bibfnamefont {A.}~\bibnamefont
  {Kitaev}},\ }\href {http://link.aps.org/doi/10.1103/PhysRevA.71.022316}
  {\bibfield  {journal} {\bibinfo  {journal} {Physical Review A}\ }\textbf
  {\bibinfo {volume} {71}},\ \bibinfo {pages} {22316} (\bibinfo {year}
  {2005})}\BibitemShut {NoStop}%
\bibitem [{\citenamefont {Li}(2015)}]{Li2015}%
  \BibitemOpen
  \bibfield  {author} {\bibinfo {author} {\bibfnamefont {Y.}~\bibnamefont
  {Li}},\ }\href {\doibase 10.1088/1367-2630/17/2/023037} {\bibfield  {journal}
  {\bibinfo  {journal} {New Journal of Physics}\ }\textbf {\bibinfo {volume}
  {17}},\ \bibinfo {pages} {5} (\bibinfo {year} {2015})}\BibitemShut {NoStop}%
\bibitem [{\citenamefont {Harness}\ and\ \citenamefont
  {Syms}(2000)}]{Harness2000}%
  \BibitemOpen
  \bibfield  {author} {\bibinfo {author} {\bibfnamefont {T.}~\bibnamefont
  {Harness}}\ and\ \bibinfo {author} {\bibfnamefont {R.~R.~A.}\ \bibnamefont
  {Syms}},\ }\href {http://stacks.iop.org/0960-1317/10/i=1/a=302} {\bibfield
  {journal} {\bibinfo  {journal} {Journal of Micromechanics and
  Microengineering}\ }\textbf {\bibinfo {volume} {10}},\ \bibinfo {pages} {7}
  (\bibinfo {year} {2000})}\BibitemShut {NoStop}%
\bibitem [{\citenamefont {Mukhopadhyay}\ \emph {et~al.}(2008)\citenamefont
  {Mukhopadhyay}, \citenamefont {Dong}, \citenamefont {Pengwang},\ and\
  \citenamefont {Ferreira}}]{Mukhopadhyay2008}%
  \BibitemOpen
  \bibfield  {author} {\bibinfo {author} {\bibfnamefont {D.}~\bibnamefont
  {Mukhopadhyay}}, \bibinfo {author} {\bibfnamefont {J.}~\bibnamefont {Dong}},
  \bibinfo {author} {\bibfnamefont {E.}~\bibnamefont {Pengwang}}, \ and\
  \bibinfo {author} {\bibfnamefont {P.}~\bibnamefont {Ferreira}},\ }\href
  {\doibase 10.1016/j.sna.2008.04.018} {\bibfield  {journal} {\bibinfo
  {journal} {Sensors and Actuators A: Physical}\ }\textbf {\bibinfo {volume}
  {147}},\ \bibinfo {pages} {340 } (\bibinfo {year} {2008})}\BibitemShut
  {NoStop}%
\bibitem [{\citenamefont {Dong}\ \emph {et~al.}(2007)\citenamefont {Dong},
  \citenamefont {Mukhopadhyay},\ and\ \citenamefont {Ferreira}}]{Dong2007}%
  \BibitemOpen
  \bibfield  {author} {\bibinfo {author} {\bibfnamefont {J.}~\bibnamefont
  {Dong}}, \bibinfo {author} {\bibfnamefont {D.}~\bibnamefont {Mukhopadhyay}},
  \ and\ \bibinfo {author} {\bibfnamefont {P.~M.}\ \bibnamefont {Ferreira}},\
  }\href {http://stacks.iop.org/0960-1317/17/i=6/a=008} {\bibfield  {journal}
  {\bibinfo  {journal} {Journal of Micromechanics and Microengineering}\
  }\textbf {\bibinfo {volume} {17}},\ \bibinfo {pages} {1154} (\bibinfo {year}
  {2007})}\BibitemShut {NoStop}%
\bibitem [{\citenamefont {Koo}\ \emph {et~al.}(2012)\citenamefont {Koo},
  \citenamefont {Zhang}, \citenamefont {Dong}, \citenamefont {Salapaka},\ and\
  \citenamefont {Ferreira}}]{Koo2012}%
  \BibitemOpen
  \bibfield  {author} {\bibinfo {author} {\bibfnamefont {B.}~\bibnamefont
  {Koo}}, \bibinfo {author} {\bibfnamefont {X.}~\bibnamefont {Zhang}}, \bibinfo
  {author} {\bibfnamefont {J.}~\bibnamefont {Dong}}, \bibinfo {author}
  {\bibfnamefont {S.}~\bibnamefont {Salapaka}}, \ and\ \bibinfo {author}
  {\bibfnamefont {P.}~\bibnamefont {Ferreira}},\ }\href {\doibase
  10.1109/JMEMS.2011.2174425} {\bibfield  {journal} {\bibinfo  {journal}
  {Journal of Microelectromechanical Systems}\ }\textbf {\bibinfo {volume}
  {21}},\ \bibinfo {pages} {13} (\bibinfo {year} {2012})}\BibitemShut {NoStop}%
\bibitem [{\citenamefont {Chu}\ and\ \citenamefont
  {Gianchandani}(2003)}]{Chu2003}%
  \BibitemOpen
  \bibfield  {author} {\bibinfo {author} {\bibfnamefont {L.~L.}\ \bibnamefont
  {Chu}}\ and\ \bibinfo {author} {\bibfnamefont {Y.~B.}\ \bibnamefont
  {Gianchandani}},\ }\href {http://stacks.iop.org/0960-1317/13/i=2/a=316}
  {\bibfield  {journal} {\bibinfo  {journal} {Journal of Micromechanics and
  Microengineering}\ }\textbf {\bibinfo {volume} {13}},\ \bibinfo {pages} {279}
  (\bibinfo {year} {2003})}\BibitemShut {NoStop}%
\bibitem [{\citenamefont {Liao}\ \emph {et~al.}(2010)\citenamefont {Liao},
  \citenamefont {Shen}, \citenamefont {Liao}, \citenamefont {Yang},
  \citenamefont {Chen},\ and\ \citenamefont {Pai}}]{Liao2010}%
  \BibitemOpen
  \bibfield  {author} {\bibinfo {author} {\bibfnamefont {H.-H.}\ \bibnamefont
  {Liao}}, \bibinfo {author} {\bibfnamefont {H.-H.}\ \bibnamefont {Shen}},
  \bibinfo {author} {\bibfnamefont {B.-T.}\ \bibnamefont {Liao}}, \bibinfo
  {author} {\bibfnamefont {Y.-J.}\ \bibnamefont {Yang}}, \bibinfo {author}
  {\bibfnamefont {Y.-C.}\ \bibnamefont {Chen}}, \ and\ \bibinfo {author}
  {\bibfnamefont {W.-W.}\ \bibnamefont {Pai}},\ }\href {\doibase
  10.1109/NEMS.2010.5592458} {\bibfield  {journal} {\bibinfo  {journal} {5th
  IEEE International Conference on Nano/Micro Engineered and Molecular Systems
  (NEMS)}\ ,\ \bibinfo {pages} {549}} (\bibinfo {year} {2010})}\BibitemShut
  {NoStop}%
\bibitem [{\citenamefont {Hile}\ \emph {et~al.}(2015)\citenamefont {Hile},
  \citenamefont {House}, \citenamefont {Peretz}, \citenamefont {Verduijn},
  \citenamefont {Widmann}, \citenamefont {Kobayashi}, \citenamefont {Rogge},\
  and\ \citenamefont {Simmons}}]{Hile2015}%
  \BibitemOpen
  \bibfield  {author} {\bibinfo {author} {\bibfnamefont {S.~J.}\ \bibnamefont
  {Hile}}, \bibinfo {author} {\bibfnamefont {M.~G.}\ \bibnamefont {House}},
  \bibinfo {author} {\bibfnamefont {E.}~\bibnamefont {Peretz}}, \bibinfo
  {author} {\bibfnamefont {J.}~\bibnamefont {Verduijn}}, \bibinfo {author}
  {\bibfnamefont {D.}~\bibnamefont {Widmann}}, \bibinfo {author} {\bibfnamefont
  {T.}~\bibnamefont {Kobayashi}}, \bibinfo {author} {\bibfnamefont
  {S.}~\bibnamefont {Rogge}}, \ and\ \bibinfo {author} {\bibfnamefont {M.~Y.}\
  \bibnamefont {Simmons}},\ }\href {\doibase
  http://dx.doi.org/10.1063/1.4929827} {\bibfield  {journal} {\bibinfo
  {journal} {Applied Physics Letters}\ }\textbf {\bibinfo {volume} {107}},\
  \bibinfo {eid} {093504} (\bibinfo {year} {2015})}\BibitemShut {NoStop}%
\bibitem [{\citenamefont {Pica}\ \emph {et~al.}(2014)\citenamefont {Pica},
  \citenamefont {Wolfowicz}, \citenamefont {Urdampilleta}, \citenamefont
  {Thewalt}, \citenamefont {Riemann}, \citenamefont {Abrosimov} \emph
  {et~al.}}]{Pica2014}%
  \BibitemOpen
  \bibfield  {author} {\bibinfo {author} {\bibfnamefont {G.}~\bibnamefont
  {Pica}}, \bibinfo {author} {\bibfnamefont {G.}~\bibnamefont {Wolfowicz}},
  \bibinfo {author} {\bibfnamefont {M.}~\bibnamefont {Urdampilleta}}, \bibinfo
  {author} {\bibfnamefont {M.~L.~W.}\ \bibnamefont {Thewalt}}, \bibinfo
  {author} {\bibfnamefont {H.}~\bibnamefont {Riemann}}, \bibinfo {author}
  {\bibfnamefont {N.~V.}\ \bibnamefont {Abrosimov}},  \emph {et~al.},\ }\href
  {\doibase 10.1103/PhysRevB.90.195204} {\bibfield  {journal} {\bibinfo
  {journal} {Phys. Rev. B}\ }\textbf {\bibinfo {volume} {90}},\ \bibinfo
  {pages} {195204} (\bibinfo {year} {2014})}\BibitemShut {NoStop}%
\bibitem [{\citenamefont {Laucht}\ \emph {et~al.}(2015)\citenamefont {Laucht},
  \citenamefont {Muhonen}, \citenamefont {Mohiyaddin}, \citenamefont {Kalra},
  \citenamefont {Dehollain}, \citenamefont {Freer}, \citenamefont {Hudson},
  \citenamefont {Veldhorst}, \citenamefont {Rahman}, \citenamefont {Klimeck},
  \citenamefont {Itoh}, \citenamefont {Jamieson}, \citenamefont {McCallum},
  \citenamefont {Dzurak},\ and\ \citenamefont {Morello}}]{Laucht2015}%
  \BibitemOpen
  \bibfield  {author} {\bibinfo {author} {\bibfnamefont {A.}~\bibnamefont
  {Laucht}}, \bibinfo {author} {\bibfnamefont {J.~T.}\ \bibnamefont {Muhonen}},
  \bibinfo {author} {\bibfnamefont {F.~A.}\ \bibnamefont {Mohiyaddin}},
  \bibinfo {author} {\bibfnamefont {R.}~\bibnamefont {Kalra}}, \bibinfo
  {author} {\bibfnamefont {J.~P.}\ \bibnamefont {Dehollain}}, \bibinfo {author}
  {\bibfnamefont {S.}~\bibnamefont {Freer}}, \bibinfo {author} {\bibfnamefont
  {F.~E.}\ \bibnamefont {Hudson}}, \bibinfo {author} {\bibfnamefont
  {M.}~\bibnamefont {Veldhorst}}, \bibinfo {author} {\bibfnamefont
  {R.}~\bibnamefont {Rahman}}, \bibinfo {author} {\bibfnamefont
  {G.}~\bibnamefont {Klimeck}}, \bibinfo {author} {\bibfnamefont {K.~M.}\
  \bibnamefont {Itoh}}, \bibinfo {author} {\bibfnamefont {D.~N.}\ \bibnamefont
  {Jamieson}}, \bibinfo {author} {\bibfnamefont {J.~C.}\ \bibnamefont
  {McCallum}}, \bibinfo {author} {\bibfnamefont {A.~S.}\ \bibnamefont
  {Dzurak}}, \ and\ \bibinfo {author} {\bibfnamefont {A.}~\bibnamefont
  {Morello}},\ }\href {\doibase 10.1126/sciadv.1500022} {\bibfield  {journal}
  {\bibinfo  {journal} {Science Advances}\ }\textbf {\bibinfo {volume} {1}}
  (\bibinfo {year} {2015}),\ 10.1126/sciadv.1500022}\BibitemShut {NoStop}%
\bibitem [{\citenamefont {Tyryshkin}\ \emph {et~al.}(2012)\citenamefont
  {Tyryshkin}, \citenamefont {Tojo}, \citenamefont {Morton}, \citenamefont
  {Riemann}, \citenamefont {Abrosimov}, \citenamefont {Becker} \emph
  {et~al.}}]{Tyryshkin2012}%
  \BibitemOpen
  \bibfield  {author} {\bibinfo {author} {\bibfnamefont {A.~M.}\ \bibnamefont
  {Tyryshkin}}, \bibinfo {author} {\bibfnamefont {S.}~\bibnamefont {Tojo}},
  \bibinfo {author} {\bibfnamefont {J.~J.~L.}\ \bibnamefont {Morton}}, \bibinfo
  {author} {\bibfnamefont {H.}~\bibnamefont {Riemann}}, \bibinfo {author}
  {\bibfnamefont {N.~V.}\ \bibnamefont {Abrosimov}}, \bibinfo {author}
  {\bibfnamefont {P.}~\bibnamefont {Becker}},  \emph {et~al.},\ }\href
  {http://dx.doi.org/10.1038/nmat3182} {\bibfield  {journal} {\bibinfo
  {journal} {Nature Materials}\ }\textbf {\bibinfo {volume} {11}},\ \bibinfo
  {pages} {143} (\bibinfo {year} {2012})}\BibitemShut {NoStop}%
\bibitem [{\citenamefont {Morello}\ \emph {et~al.}(2010)\citenamefont
  {Morello}, \citenamefont {Pla}, \citenamefont {Zwanenburg}, \citenamefont
  {Chan}, \citenamefont {Tan}, \citenamefont {Huebl} \emph
  {et~al.}}]{Morello2010}%
  \BibitemOpen
  \bibfield  {author} {\bibinfo {author} {\bibfnamefont {A.}~\bibnamefont
  {Morello}}, \bibinfo {author} {\bibfnamefont {J.~J.}\ \bibnamefont {Pla}},
  \bibinfo {author} {\bibfnamefont {F.~A.}\ \bibnamefont {Zwanenburg}},
  \bibinfo {author} {\bibfnamefont {K.~W.}\ \bibnamefont {Chan}}, \bibinfo
  {author} {\bibfnamefont {K.~Y.}\ \bibnamefont {Tan}}, \bibinfo {author}
  {\bibfnamefont {H.}~\bibnamefont {Huebl}},  \emph {et~al.},\ }\href
  {http://dx.doi.org/10.1038/nature09392} {\bibfield  {journal} {\bibinfo
  {journal} {Nature}\ }\textbf {\bibinfo {volume} {467}},\ \bibinfo {pages}
  {687} (\bibinfo {year} {2010})}\BibitemShut {NoStop}%
\bibitem [{\citenamefont {Pla}\ \emph {et~al.}(2012)\citenamefont {Pla},
  \citenamefont {Tan}, \citenamefont {Dehollain}, \citenamefont {Lim},
  \citenamefont {Morton}, \citenamefont {Jamieson} \emph {et~al.}}]{Pla2012}%
  \BibitemOpen
  \bibfield  {author} {\bibinfo {author} {\bibfnamefont {J.~J.}\ \bibnamefont
  {Pla}}, \bibinfo {author} {\bibfnamefont {K.~Y.}\ \bibnamefont {Tan}},
  \bibinfo {author} {\bibfnamefont {J.~P.}\ \bibnamefont {Dehollain}}, \bibinfo
  {author} {\bibfnamefont {W.~H.}\ \bibnamefont {Lim}}, \bibinfo {author}
  {\bibfnamefont {J.~J.~L.}\ \bibnamefont {Morton}}, \bibinfo {author}
  {\bibfnamefont {D.~N.}\ \bibnamefont {Jamieson}},  \emph {et~al.},\ }\href
  {http://dx.doi.org/10.1038/nature11449} {\bibfield  {journal} {\bibinfo
  {journal} {Nature}\ }\textbf {\bibinfo {volume} {489}},\ \bibinfo {pages}
  {541} (\bibinfo {year} {2012})}\BibitemShut {NoStop}%
\bibitem [{\citenamefont {Muhonen}\ \emph {et~al.}(2014)\citenamefont
  {Muhonen}, \citenamefont {Dehollain}, \citenamefont {Laucht}, \citenamefont
  {Hudson}, \citenamefont {Kalra}, \citenamefont {Sekiguchi} \emph
  {et~al.}}]{Muhonen2014}%
  \BibitemOpen
  \bibfield  {author} {\bibinfo {author} {\bibfnamefont {J.~T.}\ \bibnamefont
  {Muhonen}}, \bibinfo {author} {\bibfnamefont {J.~P.}\ \bibnamefont
  {Dehollain}}, \bibinfo {author} {\bibfnamefont {A.}~\bibnamefont {Laucht}},
  \bibinfo {author} {\bibfnamefont {F.~E.}\ \bibnamefont {Hudson}}, \bibinfo
  {author} {\bibfnamefont {R.}~\bibnamefont {Kalra}}, \bibinfo {author}
  {\bibfnamefont {T.}~\bibnamefont {Sekiguchi}},  \emph {et~al.},\ }\href
  {http://dx.doi.org/10.1038/nnano.2014.211} {\bibfield  {journal} {\bibinfo
  {journal} {Nat Nano}\ }\textbf {\bibinfo {volume} {9}},\ \bibinfo {pages}
  {986} (\bibinfo {year} {2014})}\BibitemShut {NoStop}%
\bibitem [{\citenamefont {Muhonen}\ \emph {et~al.}(2015)\citenamefont
  {Muhonen}, \citenamefont {Laucht}, \citenamefont {Simmons}, \citenamefont
  {Dehollain}, \citenamefont {Kalra}, \citenamefont {Hudson}, \citenamefont
  {Freer}, \citenamefont {Itoh}, \citenamefont {Jamieson}, \citenamefont
  {McCallum}, \citenamefont {Dzurak},\ and\ \citenamefont
  {Morello}}]{Muhonen2015a}%
  \BibitemOpen
  \bibfield  {author} {\bibinfo {author} {\bibfnamefont {J.~T.}\ \bibnamefont
  {Muhonen}}, \bibinfo {author} {\bibfnamefont {a.}~\bibnamefont {Laucht}},
  \bibinfo {author} {\bibfnamefont {S.}~\bibnamefont {Simmons}}, \bibinfo
  {author} {\bibfnamefont {J.~P.}\ \bibnamefont {Dehollain}}, \bibinfo {author}
  {\bibfnamefont {R.}~\bibnamefont {Kalra}}, \bibinfo {author} {\bibfnamefont
  {F.~E.}\ \bibnamefont {Hudson}}, \bibinfo {author} {\bibfnamefont
  {S.}~\bibnamefont {Freer}}, \bibinfo {author} {\bibfnamefont {K.~M.}\
  \bibnamefont {Itoh}}, \bibinfo {author} {\bibfnamefont {D.~N.}\ \bibnamefont
  {Jamieson}}, \bibinfo {author} {\bibfnamefont {J.~C.}\ \bibnamefont
  {McCallum}}, \bibinfo {author} {\bibfnamefont {A.~S.}\ \bibnamefont
  {Dzurak}}, \ and\ \bibinfo {author} {\bibfnamefont {A.}~\bibnamefont
  {Morello}},\ }\href {\doibase 10.1088/0953-8984/27/15/154205} {\bibfield
  {journal} {\bibinfo  {journal} {Journal of Physics: Condensed Matter}\
  }\textbf {\bibinfo {volume} {27}},\ \bibinfo {pages} {154205} (\bibinfo
  {year} {2015})}\BibitemShut {NoStop}%
\bibitem [{\citenamefont {Morton}\ \emph {et~al.}(2005)\citenamefont {Morton},
  \citenamefont {Tyryshkin}, \citenamefont {Ardavan}, \citenamefont
  {Porfyrakis}, \citenamefont {Lyon},\ and\ \citenamefont
  {Briggs}}]{Morton2005}%
  \BibitemOpen
  \bibfield  {author} {\bibinfo {author} {\bibfnamefont {J.~J.~L.}\
  \bibnamefont {Morton}}, \bibinfo {author} {\bibfnamefont {A.~M.}\
  \bibnamefont {Tyryshkin}}, \bibinfo {author} {\bibfnamefont {A.}~\bibnamefont
  {Ardavan}}, \bibinfo {author} {\bibfnamefont {K.}~\bibnamefont {Porfyrakis}},
  \bibinfo {author} {\bibfnamefont {S.~A.}\ \bibnamefont {Lyon}}, \ and\
  \bibinfo {author} {\bibfnamefont {G.~A.~D.}\ \bibnamefont {Briggs}},\ }\href
  {\doibase 10.1103/PhysRevLett.95.200501} {\bibfield  {journal} {\bibinfo
  {journal} {Phys. Rev. Lett.}\ }\textbf {\bibinfo {volume} {95}},\ \bibinfo
  {pages} {200501} (\bibinfo {year} {2005})}\BibitemShut {NoStop}%
\bibitem [{\citenamefont {Lo}\ \emph {et~al.}(2015)\citenamefont {Lo},
  \citenamefont {Urdampilleta}, \citenamefont {Ross}, \citenamefont
  {Gonzalez-Zalba}, \citenamefont {Mansir}, \citenamefont {Lyon}, \citenamefont
  {Thewalt},\ and\ \citenamefont {Morton}}]{Lo2015}%
  \BibitemOpen
  \bibfield  {author} {\bibinfo {author} {\bibfnamefont {C.}~\bibnamefont
  {Lo}}, \bibinfo {author} {\bibfnamefont {M.}~\bibnamefont {Urdampilleta}},
  \bibinfo {author} {\bibfnamefont {P.}~\bibnamefont {Ross}}, \bibinfo {author}
  {\bibfnamefont {M.}~\bibnamefont {Gonzalez-Zalba}}, \bibinfo {author}
  {\bibfnamefont {J.}~\bibnamefont {Mansir}}, \bibinfo {author} {\bibfnamefont
  {S.}~\bibnamefont {Lyon}}, \bibinfo {author} {\bibfnamefont {M.}~\bibnamefont
  {Thewalt}}, \ and\ \bibinfo {author} {\bibfnamefont {J.}~\bibnamefont
  {Morton}},\ }\href@noop {} {\bibfield  {journal} {\bibinfo  {journal} {Nature
  materials}\ }\textbf {\bibinfo {volume} {14}},\ \bibinfo {pages} {490}
  (\bibinfo {year} {2015})}\BibitemShut {NoStop}%
\bibitem [{Vie(2000)}]{Vieu2000}%
  \BibitemOpen
  \href {\doibase 10.1016/S0169-4332(00)00352-4} {\bibfield  {journal}
  {\bibinfo  {journal} {Applied Surface Science}\ }\textbf {\bibinfo {volume}
  {164}},\ \bibinfo {pages} {111} (\bibinfo {year} {2000})}\BibitemShut
  {NoStop}%
\bibitem [{\citenamefont {Toyli}\ \emph {et~al.}(2010)\citenamefont {Toyli},
  \citenamefont {Weis}, \citenamefont {Fuchs}, \citenamefont {Schenkel},\ and\
  \citenamefont {Awschalom}}]{Toyli2010}%
  \BibitemOpen
  \bibfield  {author} {\bibinfo {author} {\bibfnamefont {D.~M.}\ \bibnamefont
  {Toyli}}, \bibinfo {author} {\bibfnamefont {C.~D.}\ \bibnamefont {Weis}},
  \bibinfo {author} {\bibfnamefont {G.~D.}\ \bibnamefont {Fuchs}}, \bibinfo
  {author} {\bibfnamefont {T.}~\bibnamefont {Schenkel}}, \ and\ \bibinfo
  {author} {\bibfnamefont {D.~D.}\ \bibnamefont {Awschalom}},\ }\href {\doibase
  10.1021/nl102066q} {\bibfield  {journal} {\bibinfo  {journal} {Nano Letters}\
  }\textbf {\bibinfo {volume} {10}},\ \bibinfo {pages} {3168} (\bibinfo {year}
  {2010})}\BibitemShut {NoStop}%
\bibitem [{\citenamefont {Weis}\ \emph {et~al.}(2008)\citenamefont {Weis},
  \citenamefont {Schuh}, \citenamefont {Batra}, \citenamefont {Persaud},
  \citenamefont {Rangelow}, \citenamefont {Bokor} \emph {et~al.}}]{Weis2008}%
  \BibitemOpen
  \bibfield  {author} {\bibinfo {author} {\bibfnamefont {C.~D.}\ \bibnamefont
  {Weis}}, \bibinfo {author} {\bibfnamefont {A.}~\bibnamefont {Schuh}},
  \bibinfo {author} {\bibfnamefont {A.}~\bibnamefont {Batra}}, \bibinfo
  {author} {\bibfnamefont {A.}~\bibnamefont {Persaud}}, \bibinfo {author}
  {\bibfnamefont {I.~W.}\ \bibnamefont {Rangelow}}, \bibinfo {author}
  {\bibfnamefont {J.}~\bibnamefont {Bokor}},  \emph {et~al.},\ }\href {\doibase
  10.1116/1.2968614} {\bibfield  {journal} {\bibinfo  {journal} {Journal of
  Vacuum Science \& Technology B}\ }\textbf {\bibinfo {volume} {26}},\ \bibinfo
  {pages} {2596} (\bibinfo {year} {2008})}\BibitemShut {NoStop}%
\bibitem [{\citenamefont {Jamieson}\ \emph {et~al.}(2005)\citenamefont
  {Jamieson}, \citenamefont {Yang}, \citenamefont {Hopf}, \citenamefont
  {Hearne}, \citenamefont {Pakes}, \citenamefont {Prawer} \emph
  {et~al.}}]{Jamieson2005}%
  \BibitemOpen
  \bibfield  {author} {\bibinfo {author} {\bibfnamefont {D.~N.}\ \bibnamefont
  {Jamieson}}, \bibinfo {author} {\bibfnamefont {C.}~\bibnamefont {Yang}},
  \bibinfo {author} {\bibfnamefont {T.}~\bibnamefont {Hopf}}, \bibinfo {author}
  {\bibfnamefont {S.~M.}\ \bibnamefont {Hearne}}, \bibinfo {author}
  {\bibfnamefont {C.~I.}\ \bibnamefont {Pakes}}, \bibinfo {author}
  {\bibfnamefont {S.}~\bibnamefont {Prawer}},  \emph {et~al.},\ }\href
  {\doibase 10.1063/1.1925320} {\bibfield  {journal} {\bibinfo  {journal}
  {Applied Physics Letters}\ }\textbf {\bibinfo {volume} {86}},\ \bibinfo {eid}
  {202101} (\bibinfo {year} {2005})}\BibitemShut {NoStop}%
\bibitem [{\citenamefont {Robledo}\ \emph {et~al.}(2011)\citenamefont
  {Robledo}, \citenamefont {Childress}, \citenamefont {Bernien}, \citenamefont
  {Hensen}, \citenamefont {Alkemade},\ and\ \citenamefont
  {Hanson}}]{Robledo2011}%
  \BibitemOpen
  \bibfield  {author} {\bibinfo {author} {\bibfnamefont {L.}~\bibnamefont
  {Robledo}}, \bibinfo {author} {\bibfnamefont {L.}~\bibnamefont {Childress}},
  \bibinfo {author} {\bibfnamefont {H.}~\bibnamefont {Bernien}}, \bibinfo
  {author} {\bibfnamefont {B.}~\bibnamefont {Hensen}}, \bibinfo {author}
  {\bibfnamefont {P.~F.~A.}\ \bibnamefont {Alkemade}}, \ and\ \bibinfo {author}
  {\bibfnamefont {R.}~\bibnamefont {Hanson}},\ }\href
  {http://dx.doi.org/10.1038/nature10401} {\bibfield  {journal} {\bibinfo
  {journal} {Nature}\ }\textbf {\bibinfo {volume} {477}},\ \bibinfo {pages}
  {574} (\bibinfo {year} {2011})}\BibitemShut {NoStop}%
\bibitem [{\citenamefont {Pfaff}\ \emph {et~al.}(2014)\citenamefont {Pfaff},
  \citenamefont {Hensen}, \citenamefont {Bernien}, \citenamefont {van Dam},
  \citenamefont {Blok}, \citenamefont {Taminiau}, \citenamefont {Tiggelman},
  \citenamefont {Schouten}, \citenamefont {Markham}, \citenamefont {Twitchen},\
  and\ \citenamefont {Hanson}}]{Pfaff2014}%
  \BibitemOpen
  \bibfield  {author} {\bibinfo {author} {\bibfnamefont {W.}~\bibnamefont
  {Pfaff}}, \bibinfo {author} {\bibfnamefont {B.~J.}\ \bibnamefont {Hensen}},
  \bibinfo {author} {\bibfnamefont {H.}~\bibnamefont {Bernien}}, \bibinfo
  {author} {\bibfnamefont {S.~B.}\ \bibnamefont {van Dam}}, \bibinfo {author}
  {\bibfnamefont {M.~S.}\ \bibnamefont {Blok}}, \bibinfo {author}
  {\bibfnamefont {T.~H.}\ \bibnamefont {Taminiau}}, \bibinfo {author}
  {\bibfnamefont {M.~J.}\ \bibnamefont {Tiggelman}}, \bibinfo {author}
  {\bibfnamefont {R.~N.}\ \bibnamefont {Schouten}}, \bibinfo {author}
  {\bibfnamefont {M.}~\bibnamefont {Markham}}, \bibinfo {author} {\bibfnamefont
  {D.~J.}\ \bibnamefont {Twitchen}}, \ and\ \bibinfo {author} {\bibfnamefont
  {R.}~\bibnamefont {Hanson}},\ }\href {\doibase 10.1126/science.1253512}
  {\bibfield  {journal} {\bibinfo  {journal} {Science}\ }\textbf {\bibinfo
  {volume} {345}},\ \bibinfo {pages} {532} (\bibinfo {year}
  {2014})}\BibitemShut {NoStop}%
\bibitem [{\citenamefont {Bar-Gill}\ \emph {et~al.}(2013)\citenamefont
  {Bar-Gill}, \citenamefont {Pham}, \citenamefont {Jarmola}, \citenamefont
  {Budker},\ and\ \citenamefont {Walsworth}}]{BarGill2013}%
  \BibitemOpen
  \bibfield  {author} {\bibinfo {author} {\bibfnamefont {N.}~\bibnamefont
  {Bar-Gill}}, \bibinfo {author} {\bibfnamefont {L.~M.}\ \bibnamefont {Pham}},
  \bibinfo {author} {\bibfnamefont {A.}~\bibnamefont {Jarmola}}, \bibinfo
  {author} {\bibfnamefont {D.}~\bibnamefont {Budker}}, \ and\ \bibinfo {author}
  {\bibfnamefont {R.~L.}\ \bibnamefont {Walsworth}},\ }\href
  {http://dx.doi.org/10.1038/ncomms2771} {\bibfield  {journal} {\bibinfo
  {journal} {Nat Commun}\ }\textbf {\bibinfo {volume} {4}},\ \bibinfo {pages}
  {1743} (\bibinfo {year} {2013})}\BibitemShut {NoStop}%
\bibitem [{\citenamefont {Dolde}\ \emph {et~al.}(2011)\citenamefont {Dolde},
  \citenamefont {Fedder}, \citenamefont {Doherty}, \citenamefont {N{\"o}bauer},
  \citenamefont {Rempp}, \citenamefont {Balasubramanian}, \citenamefont {Wolf},
  \citenamefont {Reinhard}, \citenamefont {Hollenberg}, \citenamefont {Jelezko}
  \emph {et~al.}}]{Dolde2011}%
  \BibitemOpen
  \bibfield  {author} {\bibinfo {author} {\bibfnamefont {F.}~\bibnamefont
  {Dolde}}, \bibinfo {author} {\bibfnamefont {H.}~\bibnamefont {Fedder}},
  \bibinfo {author} {\bibfnamefont {M.~W.}\ \bibnamefont {Doherty}}, \bibinfo
  {author} {\bibfnamefont {T.}~\bibnamefont {N{\"o}bauer}}, \bibinfo {author}
  {\bibfnamefont {F.}~\bibnamefont {Rempp}}, \bibinfo {author} {\bibfnamefont
  {G.}~\bibnamefont {Balasubramanian}}, \bibinfo {author} {\bibfnamefont
  {T.}~\bibnamefont {Wolf}}, \bibinfo {author} {\bibfnamefont {F.}~\bibnamefont
  {Reinhard}}, \bibinfo {author} {\bibfnamefont {L.}~\bibnamefont
  {Hollenberg}}, \bibinfo {author} {\bibfnamefont {F.}~\bibnamefont {Jelezko}},
   \emph {et~al.},\ }\href@noop {} {\bibfield  {journal} {\bibinfo  {journal}
  {Nature Physics}\ }\textbf {\bibinfo {volume} {7}},\ \bibinfo {pages} {459}
  (\bibinfo {year} {2011})}\BibitemShut {NoStop}%
\bibitem [{\citenamefont {Ovartchaiyapong}\ \emph {et~al.}(2012)\citenamefont
  {Ovartchaiyapong}, \citenamefont {Pascal}, \citenamefont {Myers},
  \citenamefont {Lauria},\ and\ \citenamefont
  {Bleszynski~Jayich}}]{Ovartchaiyapong2012}%
  \BibitemOpen
  \bibfield  {author} {\bibinfo {author} {\bibfnamefont {P.}~\bibnamefont
  {Ovartchaiyapong}}, \bibinfo {author} {\bibfnamefont {L.~M.~A.}\ \bibnamefont
  {Pascal}}, \bibinfo {author} {\bibfnamefont {B.~A.}\ \bibnamefont {Myers}},
  \bibinfo {author} {\bibfnamefont {P.}~\bibnamefont {Lauria}}, \ and\ \bibinfo
  {author} {\bibfnamefont {A.~C.}\ \bibnamefont {Bleszynski~Jayich}},\ }\href
  {\doibase 10.1063/1.4760274} {\bibfield  {journal} {\bibinfo  {journal}
  {Applied Physics Letters}\ }\textbf {\bibinfo {volume} {101}},\ \bibinfo
  {eid} {163505} (\bibinfo {year} {2012})}\BibitemShut {NoStop}%
\bibitem [{\citenamefont {Imboden}\ \emph {et~al.}(2013)\citenamefont
  {Imboden}, \citenamefont {Williams},\ and\ \citenamefont
  {Mohanty}}]{Imboden2013}%
  \BibitemOpen
  \bibfield  {author} {\bibinfo {author} {\bibfnamefont {M.}~\bibnamefont
  {Imboden}}, \bibinfo {author} {\bibfnamefont {O.~A.}\ \bibnamefont
  {Williams}}, \ and\ \bibinfo {author} {\bibfnamefont {P.}~\bibnamefont
  {Mohanty}},\ }\href {\doibase 10.1021/nl401978p} {\bibfield  {journal}
  {\bibinfo  {journal} {Nano Letters}\ }\textbf {\bibinfo {volume} {13}},\
  \bibinfo {pages} {4014} (\bibinfo {year} {2013})}\BibitemShut {NoStop}%
\bibitem [{\citenamefont {Williams}(2011)}]{Williams2011}%
  \BibitemOpen
  \bibfield  {author} {\bibinfo {author} {\bibfnamefont {O.~A.}\ \bibnamefont
  {Williams}},\ }\href {\doibase 10.1016/j.diamond.2011.02.015} {\bibfield
  {journal} {\bibinfo  {journal} {Diamond and Related Materials}\ }\textbf
  {\bibinfo {volume} {20}},\ \bibinfo {pages} {621 } (\bibinfo {year}
  {2011})}\BibitemShut {NoStop}%
\bibitem [{\citenamefont {Pezzagna}\ \emph {et~al.}(2010)\citenamefont
  {Pezzagna}, \citenamefont {Wildanger}, \citenamefont {Mazarov}, \citenamefont
  {Wieck}, \citenamefont {Sarov}, \citenamefont {Rangelow} \emph
  {et~al.}}]{Pezzagna2010}%
  \BibitemOpen
  \bibfield  {author} {\bibinfo {author} {\bibfnamefont {S.}~\bibnamefont
  {Pezzagna}}, \bibinfo {author} {\bibfnamefont {D.}~\bibnamefont {Wildanger}},
  \bibinfo {author} {\bibfnamefont {P.}~\bibnamefont {Mazarov}}, \bibinfo
  {author} {\bibfnamefont {A.~D.}\ \bibnamefont {Wieck}}, \bibinfo {author}
  {\bibfnamefont {Y.}~\bibnamefont {Sarov}}, \bibinfo {author} {\bibfnamefont
  {I.}~\bibnamefont {Rangelow}},  \emph {et~al.},\ }\href {\doibase
  10.1002/smll.201000902} {\bibfield  {journal} {\bibinfo  {journal} {Small}\
  }\textbf {\bibinfo {volume} {6}},\ \bibinfo {pages} {2117} (\bibinfo {year}
  {2010})}\BibitemShut {NoStop}%
\bibitem [{\citenamefont {Lesik}\ \emph {et~al.}(2013)\citenamefont {Lesik},
  \citenamefont {Spinicelli}, \citenamefont {Pezzagna}, \citenamefont {Happel},
  \citenamefont {Jacques}, \citenamefont {Salord} \emph {et~al.}}]{Lesik2013}%
  \BibitemOpen
  \bibfield  {author} {\bibinfo {author} {\bibfnamefont {M.}~\bibnamefont
  {Lesik}}, \bibinfo {author} {\bibfnamefont {P.}~\bibnamefont {Spinicelli}},
  \bibinfo {author} {\bibfnamefont {S.}~\bibnamefont {Pezzagna}}, \bibinfo
  {author} {\bibfnamefont {P.}~\bibnamefont {Happel}}, \bibinfo {author}
  {\bibfnamefont {V.}~\bibnamefont {Jacques}}, \bibinfo {author} {\bibfnamefont
  {O.}~\bibnamefont {Salord}},  \emph {et~al.},\ }\href {\doibase
  10.1002/pssa.201300102} {\bibfield  {journal} {\bibinfo  {journal} {physica
  status solidi (a)}\ }\textbf {\bibinfo {volume} {210}},\ \bibinfo {pages}
  {2055} (\bibinfo {year} {2013})}\BibitemShut {NoStop}%
\bibitem [{\citenamefont {Naydenov}\ \emph {et~al.}(2010)\citenamefont
  {Naydenov}, \citenamefont {Richter}, \citenamefont {Beck}, \citenamefont
  {Steiner}, \citenamefont {Neumann}, \citenamefont {Balasubramanian} \emph
  {et~al.}}]{Naydenov2010}%
  \BibitemOpen
  \bibfield  {author} {\bibinfo {author} {\bibfnamefont {B.}~\bibnamefont
  {Naydenov}}, \bibinfo {author} {\bibfnamefont {V.}~\bibnamefont {Richter}},
  \bibinfo {author} {\bibfnamefont {J.}~\bibnamefont {Beck}}, \bibinfo {author}
  {\bibfnamefont {M.}~\bibnamefont {Steiner}}, \bibinfo {author} {\bibfnamefont
  {P.}~\bibnamefont {Neumann}}, \bibinfo {author} {\bibfnamefont
  {G.}~\bibnamefont {Balasubramanian}},  \emph {et~al.},\ }\href {\doibase
  10.1063/1.3409221} {\bibfield  {journal} {\bibinfo  {journal} {Applied
  Physics Letters}\ }\textbf {\bibinfo {volume} {96}},\ \bibinfo {eid} {163108}
  (\bibinfo {year} {2010})}\BibitemShut {NoStop}%
\bibitem [{\citenamefont {Grinolds}\ \emph {et~al.}(2013)\citenamefont
  {Grinolds}, \citenamefont {Hong}, \citenamefont {Maletinsky}, \citenamefont
  {Luan}, \citenamefont {Lukin}, \citenamefont {Walsworth} \emph
  {et~al.}}]{Grinolds2013}%
  \BibitemOpen
  \bibfield  {author} {\bibinfo {author} {\bibfnamefont {M.~S.}\ \bibnamefont
  {Grinolds}}, \bibinfo {author} {\bibfnamefont {S.}~\bibnamefont {Hong}},
  \bibinfo {author} {\bibfnamefont {P.}~\bibnamefont {Maletinsky}}, \bibinfo
  {author} {\bibfnamefont {L.}~\bibnamefont {Luan}}, \bibinfo {author}
  {\bibfnamefont {M.~D.}\ \bibnamefont {Lukin}}, \bibinfo {author}
  {\bibfnamefont {R.~L.}\ \bibnamefont {Walsworth}},  \emph {et~al.},\ }\href
  {http://dx.doi.org/10.1038/nphys2543} {\bibfield  {journal} {\bibinfo
  {journal} {Nat Phys}\ }\textbf {\bibinfo {volume} {9}},\ \bibinfo {pages}
  {215} (\bibinfo {year} {2013})}\BibitemShut {NoStop}%
\bibitem [{\citenamefont {Koehl}\ \emph {et~al.}(2011)\citenamefont {Koehl},
  \citenamefont {Buckley}, \citenamefont {Heremans}, \citenamefont {Calusine},\
  and\ \citenamefont {Awschalom}}]{Koehl2011}%
  \BibitemOpen
  \bibfield  {author} {\bibinfo {author} {\bibfnamefont {W.~F.}\ \bibnamefont
  {Koehl}}, \bibinfo {author} {\bibfnamefont {B.~B.}\ \bibnamefont {Buckley}},
  \bibinfo {author} {\bibfnamefont {F.~J.}\ \bibnamefont {Heremans}}, \bibinfo
  {author} {\bibfnamefont {G.}~\bibnamefont {Calusine}}, \ and\ \bibinfo
  {author} {\bibfnamefont {D.~D.}\ \bibnamefont {Awschalom}},\ }\href
  {http://dx.doi.org/10.1038/nature10562} {\bibfield  {journal} {\bibinfo
  {journal} {Nature}\ }\textbf {\bibinfo {volume} {479}},\ \bibinfo {pages}
  {84} (\bibinfo {year} {2011})}\BibitemShut {NoStop}%
\bibitem [{\citenamefont {Adachi}\ \emph {et~al.}(2013)\citenamefont {Adachi},
  \citenamefont {Watanabe}, \citenamefont {Okamoto}, \citenamefont {Yamaguchi},
  \citenamefont {Kimoto},\ and\ \citenamefont {Suda}}]{Adachi2013}%
  \BibitemOpen
  \bibfield  {author} {\bibinfo {author} {\bibfnamefont {K.}~\bibnamefont
  {Adachi}}, \bibinfo {author} {\bibfnamefont {N.}~\bibnamefont {Watanabe}},
  \bibinfo {author} {\bibfnamefont {H.}~\bibnamefont {Okamoto}}, \bibinfo
  {author} {\bibfnamefont {H.}~\bibnamefont {Yamaguchi}}, \bibinfo {author}
  {\bibfnamefont {T.}~\bibnamefont {Kimoto}}, \ and\ \bibinfo {author}
  {\bibfnamefont {J.}~\bibnamefont {Suda}},\ }\href {\doibase
  10.1016/j.sna.2013.04.014} {\bibfield  {journal} {\bibinfo  {journal}
  {Sensors and Actuators A: Physical}\ }\textbf {\bibinfo {volume} {197}},\
  \bibinfo {pages} {122 } (\bibinfo {year} {2013})}\BibitemShut {NoStop}%
\bibitem [{\citenamefont {Christle}\ \emph {et~al.}(2015)\citenamefont
  {Christle}, \citenamefont {Falk}, \citenamefont {Andrich}, \citenamefont
  {Klimov}, \citenamefont {Hassan}, \citenamefont {Son}, \citenamefont
  {Janz{\'e}n} \emph {et~al.}}]{Christle2014}%
  \BibitemOpen
  \bibfield  {author} {\bibinfo {author} {\bibfnamefont {D.~J.}\ \bibnamefont
  {Christle}}, \bibinfo {author} {\bibfnamefont {A.~L.}\ \bibnamefont {Falk}},
  \bibinfo {author} {\bibfnamefont {P.}~\bibnamefont {Andrich}}, \bibinfo
  {author} {\bibfnamefont {P.~V.}\ \bibnamefont {Klimov}}, \bibinfo {author}
  {\bibfnamefont {J.~U.}\ \bibnamefont {Hassan}}, \bibinfo {author}
  {\bibfnamefont {N.~T.}\ \bibnamefont {Son}}, \bibinfo {author} {\bibfnamefont
  {E.}~\bibnamefont {Janz{\'e}n}},  \emph {et~al.},\ }\href
  {http://dx.doi.org/10.1038/nmat4144} {\bibfield  {journal} {\bibinfo
  {journal} {Nat Mater}\ }\textbf {\bibinfo {volume} {14}},\ \bibinfo {pages}
  {160} (\bibinfo {year} {2015})}\BibitemShut {NoStop}%
\bibitem [{\citenamefont {Morton}\ \emph {et~al.}(2008)\citenamefont {Morton},
  \citenamefont {Tyryshkin}, \citenamefont {Brown}, \citenamefont {Shankar},
  \citenamefont {Lovett}, \citenamefont {Ardavan} \emph {et~al.}}]{Morton2008}%
  \BibitemOpen
  \bibfield  {author} {\bibinfo {author} {\bibfnamefont {J.~J.~L.}\
  \bibnamefont {Morton}}, \bibinfo {author} {\bibfnamefont {A.~M.}\
  \bibnamefont {Tyryshkin}}, \bibinfo {author} {\bibfnamefont {R.~M.}\
  \bibnamefont {Brown}}, \bibinfo {author} {\bibfnamefont {S.}~\bibnamefont
  {Shankar}}, \bibinfo {author} {\bibfnamefont {B.~W.}\ \bibnamefont {Lovett}},
  \bibinfo {author} {\bibfnamefont {A.}~\bibnamefont {Ardavan}},  \emph
  {et~al.},\ }\href {http://dx.doi.org/10.1038/nature07295} {\bibfield
  {journal} {\bibinfo  {journal} {Nature}\ }\textbf {\bibinfo {volume} {455}},\
  \bibinfo {pages} {1085} (\bibinfo {year} {2008})}\BibitemShut {NoStop}%
\bibitem [{\citenamefont {Waldherr}\ \emph {et~al.}(2014)\citenamefont
  {Waldherr}, \citenamefont {Wang}, \citenamefont {Zaiser}, \citenamefont
  {Jamali}, \citenamefont {Schulte-Herbruggen}, \citenamefont {Abe} \emph
  {et~al.}}]{Waldherr2014}%
  \BibitemOpen
  \bibfield  {author} {\bibinfo {author} {\bibfnamefont {G.}~\bibnamefont
  {Waldherr}}, \bibinfo {author} {\bibfnamefont {Y.}~\bibnamefont {Wang}},
  \bibinfo {author} {\bibfnamefont {S.}~\bibnamefont {Zaiser}}, \bibinfo
  {author} {\bibfnamefont {M.}~\bibnamefont {Jamali}}, \bibinfo {author}
  {\bibfnamefont {T.}~\bibnamefont {Schulte-Herbruggen}}, \bibinfo {author}
  {\bibfnamefont {H.}~\bibnamefont {Abe}},  \emph {et~al.},\ }\href
  {http://dx.doi.org/10.1038/nature12919} {\bibfield  {journal} {\bibinfo
  {journal} {Nature}\ }\textbf {\bibinfo {volume} {506}},\ \bibinfo {pages}
  {204} (\bibinfo {year} {2014})}\BibitemShut {NoStop}%
\bibitem [{\citenamefont {Fowler}\ and\ \citenamefont
  {Martinis}(2014)}]{FowlerCorrelated}%
  \BibitemOpen
  \bibfield  {author} {\bibinfo {author} {\bibfnamefont {A.~G.}\ \bibnamefont
  {Fowler}}\ and\ \bibinfo {author} {\bibfnamefont {J.~M.}\ \bibnamefont
  {Martinis}},\ }\href {http://link.aps.org/doi/10.1103/PhysRevA.89.032316}
  {\bibfield  {journal} {\bibinfo  {journal} {Physical Review A}\ }\textbf
  {\bibinfo {volume} {89}},\ \bibinfo {pages} {32316} (\bibinfo {year}
  {2014})}\BibitemShut {NoStop}%
\end{thebibliography}%

\bigskip

\onecolumngrid

\bigskip
\bigskip
\bigskip

\pagebreak
\centerline{ \bf \Large Appendices}

 
 
 \section{Error Model \label{Appendix:ErrorModel}}
Aside from the permanent misalignment in the physical location of the probe and data qubits, which we discuss in detail in the main paper, we must also account for random errors in the preparation, control and measurement of our spin states. Each has an associated noise model. 

 \begin{enumerate}

\item Preparation of the probe state. This is modelled by assuming the probe is prepared in the ideal state and then with probability $p_{prep}$ this state is subjected to a randomly selected Pauli $X$, $Y$ or $Z$ rotation. However since our ideal state $\ket{+}$ is an eigenstate of $X$, only the latter two operations will be significant.

\item Controlled rotation of a spin (e.g. either a flip of the probe spin as part of the `twirl' process, or a Hadamard operation on a data qubit).
With probability $p_{single}$ the ideal operation is followed by a randomly selected Pauli $X$, $Y$ or $Z$ rotation. 

\item `Jitter', i.e. random variations in the interaction between the probe and a data qubit, as opposed to the systematic errors due to permanent misalignment. We assume that the actual phase acquired in the two-qubit operation $S$ \textcolor{black}{varies randomly and uniformly between limits $\frac{\pi}{2}\pm \phi_e$, and in our simulations we found we could set $\phi_e=(0.044)\frac{\pi}{2}$, i.e. $4.4\%$ of the ideal phase, before there was any appreciable impact on the threshold. We therefore selected this level of jitter, as reported in the table within Fig.~\ref{fig:errorTable}. The reason for this remarkable level of tolerance is a quadratic relation between the unwanted phase shift (which is proportional to physical imperfections such as, e.g., a timing error) and the actual probability of a discretised error entering the model. The following text summaries the argument.}

\textcolor{black}{Consider first the simple bimodal case where phase $\frac{\pi}{2}+\phi_b$ occurs with probability one half, and  $\frac{\pi}{2}-\phi_b$ occurs with probability one half. The evolution of the quantum state can therefore be written as the ideal phase gate $S(\frac{\pi}{2})$ followed by an error mapping $\rho$ of the form
\begin{eqnarray}
\rho&\rightarrow&\frac{1}{2}S(\phi_b)\,\rho\, S^\dagger(\phi_b)+\frac{1}{2}S(-\phi_b)\,\rho\, S^\dagger(-\phi_b)\nonumber\\
&=&\left(\cos^2\frac{\phi_b}{2}\right)\rho+\left(\sin^2\frac{\phi_b}{2}\right)Z_1Z_2\,\rho\, Z_1 Z_2
\end{eqnarray}
where $Z_1$ and $Z_2$ are Pauli operators on the probe and data qubit as in Eqn.~\ref{eqn:S}. This is therefore equivalent to discrete error event $Z_1Z_2$ occurring with probability $\sin^2(\phi_b/2)$ or approximately $\phi_b^2/4$ for small $\phi_b$.}

\textcolor{black}{If we now consider any symmetric distribution of possible phase errors, i.e. a probability density $p(\phi)$ of an unwanted phase shift between $\phi$ and $\phi+d\phi$ where $p(-\phi)=p(\phi)$, then can can simply match positive and negative shifts as above and integrate, so that our mapping is $\rho\rightarrow (1-\epsilon)\rho+\epsilon Z_1Z_2\,\rho\, Z_1Z_2$ with
\[
\epsilon=2\int_0^\infty p(x)\sin^2(\frac{x}{2})\,dx.
\]
Taking our uniform distribution of phase error from $+\phi_e$ to $-\phi_e$ one finds that $\epsilon\simeq \phi_e^2/12$ for small $\phi_e$. With our choice of a $4.4\%$ jitter, i.e. $\phi_e=(0.044)\frac{\pi}{2}$, this corresponds to $\epsilon=4\times10^{-4}$ i.e. a very small $0.04\%$ probability of the $Z_1Z_2$ error event.}

\item Measurement. We select a measurement error rate $p_m$ and then a particular outcome of the measurement, $q\in\left\{ 0,1\right\} $ corresponds to the intended projection
$P_{q}$ applied to the state with probability $(1-p_{m})$ and the opposite projection $P_{\bar{q}}$ applied with probability $p_{m}$. This noisy projector can be written: 
\begin{equation}
\mathcal{P}_{q}\left(p_{m}\right)=\left(1-p_{m}\right)|q\rangle\langle q|+p_{m}|\bar{q}\rangle\langle\bar{q}|
\end{equation}
In a refinement of this model, we can enter two different values of $p_m$, one for the cause that $j=0$ and one for $j=1$. This reflects the reality of many experimental realisations of measurement where, e.g.,  $\ket{1}$ is associated with an active detection event and $\ket{0}$ is associated with that event not occurring (in optical measurement, the event is seeing a photon that is characteristic of $\ket{1}$). Because of the asymmetry of the process, once imperfections such as photon loss are allowed for then the fidelity of measurement becomes dependent on the state that is measured, $\ket{0}$ or $\ket{1}$.

\item \textcolor{black}{ Data qubit error. We model decoherence of the data qubits that occurs during the timescale of a stabilizer cycle. At the end of each round of stabilizer measurements each data qubit is subjected to a random Pauli $X$, $Y$ or $Z$ error with probability $p_{data}$.}

\end{enumerate}

\subsection{\textcolor{black}{Discretisation of errors}}
\textcolor{black}{Our numerical simulations use the standard approach: we discretise errors into Pauli events that either do or do not occur, with some given probability. Thresholds found in this way should accurately reflect the performance of a machine in which errors have a more general form, including small coherent imperfections.}

\textcolor{black}{If two errors $E_1$ and $E_2$ are correctable with a certain code then every linear combination of these $\alpha E_1 + \beta E_2$ with $\alpha, \beta \in \mathbb{C}$ is also correctable. The act of making a syndrome measurement will project into the state where either $E_1$ or $E_2$ occurred, and these errors are by definition correctable. As the Pauli operators form the basis of operators for an $n$-qubit Hilbert space, the ability to correct Pauli errors of a certain weight allows us to correct all errors up to that weight. }

\textcolor{black}{Errors associated with our active manipulations (initialisation, periods of interaction, control pulses, measurements) are modelled as occuring at the time of that manipulation. Meanwhile it is convenient to model `environmental decoherence' on our data qubits, i.e. errors that are not associated with our active manipulations, by applying Pauli errors at the end of each round of stabilizer measurements. In reality such effects can occur at any time; however the consequences of any change to the state of a data qubit arise only once that qubit interacts with a probe.
Introducing errors discretely at a fixed time effectively accounts for the accumulation of error probability over the time the data qubit is isolated (and this is the majority of the time: $\sim 98\%$ for the circular orbit model). There will be a some small decoherence effect during the probe-data qubit interaction, but even this is approximately captured by our model. To see this we need only consider $X$-type errors since $Z$ errors have no immediate effect -- they commute with the interaction $S$, see Eqn.~\ref{eqn:S}. Then note that such an error gives rise to a superposition of two states equivalent to ``flip suffered immediately before the interaction'' and ``flip suffered immediately after the interaction''; subsequent measurement of the probe collapses this superposition since the terms lead to different parity measurements.}

 \section{stabilizers as superoperator\label{Appendix:AnalyticSO}}

To characterise the entire process of the stabilizer measurement we carry out a full analysis of the measurement procedure including all sources of noise noted in the section above, and generate a superoperator from the result to completely describe the action of the stabilizing measurement procedure. 

\begin{equation}
\mathcal{S}\left(\rho\right)=\sum_{i=0}p_{i}K_{i}\rho K_{i}^{\dagger}
\end{equation}

This probabilistic decomposition describes the operation as a series of Kraus operators, $K_{i}$, applied to the initial state with probabilities $p_{i}$, which depend on the chosen protocol, noise model and the error rates. The leading term $i=0$ will have corresponding $K_{0}$ representing the reported parity projection, and large $p_{0}$. For the protocols considered here, the other Kraus operations can be decomposed and expressed as a parity projection with additional erroneous operations applied.

Consider a known deterministic set of phase errors over a 4-qubit stabilizer-by-probe. The probe and data qubits mutually acquire phase through their dipole-dipole interaction. This interaction between probe and single data qubit leads to the following gate

\begin{equation}
\label{eqn::perfect}
S(\theta)=\left(\begin{array}{cccc}
1&0&0&0\\
0&\exp(i\theta)&0&0\\
0&0&\exp(i\theta)&0\\
0&0&0&1
\end{array}\right)
\end{equation}
where $\theta = \pi/2 + \delta(x,y,z)$ is a function of the position of the data qubit. 

This means that after the probe has passed over one of the four qubits the state of the system is
\begin{equation}
\label{eqn::state after one interaction}
\left|system\right> \propto \left(\left|0\right>V^a S^a_1 + \left|1\right> (i V^a Z^a) S^{'a}_1\right)\left|data\right>,
\end{equation}
where
$V=diag\{1,i\} $, $S_k=diag\{1,e^{i \delta_k}\}$, $S^{'}_k=diag\{e^{i\delta_k},1\}$, $Z=diag\{1,-1\}$ and the superscripts $\{a,b,c,d\}$ label the data qubit on which the operator acts.

After the probe has passed four data qubits, each of which injects some erroneous phase $\delta_i$ onto the probe qubit, the state of the system is proportional to
\begin{equation}
\label{eqn::after four interaction}
\left(\left|0\right> S^d_4S^c_3S^b_2S^a_1 + \left|1\right> Z^d Z^c Z^b Z^a S^{'d}_4 S^{'c}_3 S^{'b}_2 S^{'a}_1 \right)V^d V^c V^b V^a \left|data\right>. 
\end{equation}

We want to measure the probe in the $\left|\pm\right>$ basis, the result of which will determine our estimate of the parity of the data qubits. Rewriting Equation \ref{eqn::after four interaction},
\begin{equation}
\label{eqn::premeasure state}
\left|\pm \right>\left( S^d_4S^c_3S^b_2S^a_1 \pm Z^d Z^c Z^b Z^a S^{'d}_4 S^{'c}_3 S^{'b}_2 S^{'a}_1 \right)V^d V^c V^b V^a \left|data\right>.
\end{equation}

If measurement of the probe finds it in the $\ket{+}$ then we interpret this as an attempted even parity projection. We neglect for now the unconditional phases $V^d V^c V^b V ^a$. The actual projection we have performed on the data qubits is

\begin{equation}
\label{eqn::wonky}
\begin{aligned}
\hat{P}^{\prime}_\text{even}&=  \ S^d_4 S^c_3 S^b_2 S^a_1 + Z^d Z^c Z^b  Z^a S^{'d}_4 S^{'c}_3 S^{'b}_2 S^{'a}_1 \\
&\propto c_1\left(\left|0000\right>\left<0000\right|+\left|1111\right>\left<1111 \right| \right) +\\
&\ \ \ \ c_2 \left(\ket{0011}\left<0011\right|+\left|1100\right>\left<1100\right|\right) +\\
&\ \ \ \ c_3 \left(\left|0101\right>\left<0101\right|+\left|1010\right>\left<1010\right|\right) +\\
&\ \ \ \ c_4 \left(\left|0110\right>\left<0110\right|+\left|1001\right>\left<1001\right|\right) +\\
&\ \ \ i s_1\left(\left|1110\right>\left<1110\right|-\left|0001\right>\left<0001\right|\right)  +\\
&\ \ \ i s_2\left(\left|1101\right>\left<1101\right|-\left|0010\right>\left<0010\right|\right) +\\
&\ \ \ i s_3\left(\left|1011\right>\left<1011\right|-\left|0100\right>\left<0100\right|\right)  +\\
&\ \ \ i s_4\left(\left|1000\right>\left<1000\right|-\left|0111\right>\left<0111\right|\right).
\end{aligned}
\end{equation}
This is clearly not a true parity projection, as different even parity subspaces $P^{(i)}_{\text{even}}$ have different weightings $c_i$, e.g. $c_1=\cos\left(\frac{\delta_1+\delta_2+\delta_3+\delta_4}{2}\right)$ for $P^{(1)}_{\text{even}}=\ket{0000}+\ket{1111}$ and there is some weight on projection onto odd parity subspaces $P^{(i)}_{\text{odd}}$, e.g. $s_1=\sin\left(\frac{\delta_1+\delta_2+\delta_3-\delta_4}{2}\right)$ for $P^{(1)}_{\text{odd}}=\ket{1110}+\ket{0001}$.

Consider the following protocol to smooth out this systematic error in our parity measurement: we randomly select one of four patterns of X operators on the data qubits and apply it before and after the $\hat{P}^{\prime}_\text{even}$ projector. We choose from the set $U_1=\mathbb{1}\mathbb{1}\mathbb{1}\mathbb{1}, U_2=\mathbb{1}\mathbb{1}XX, U_3=\mathbb{1}X\mathbb{1}X,U_4=\mathbb{1}XX\mathbb{1}$ to smooth the weightings $c_i$ and $s_i$ of Equation \ref{eqn::wonky}. 

The action of this protocol on the state $\rho$ of the data qubits is thus,
\begin{equation}
\label{eqn::Xflip protocol}
\begin{aligned}
P^{\text{smooth}}_{\text{even}}(\rho)=  \frac{1}{4} \big[
&(U_1\hat{P}^{\prime}_\text{even}U_1)\rho (U_1 \hat{P}^{\prime \dagger}_{\text{even}} U_1) +\\
&(U_2\hat{P}^{\prime}_\text{even}U_2)\rho (U_2 \hat{P}^{\prime \dagger}_{\text{even}} U_2) +\\
&(U_3\hat{P}^{\prime}_\text{even}U_3)\rho (U_3 \hat{P}^{\prime \dagger}_{\text{even}} U_3) +\\
&(U_4\hat{P}^{\prime}_\text{even}U_4)\rho (U_4 \hat{P}^{\prime \dagger}_{\text{even}} U_4) \big]
\end{aligned}.
\end{equation}
The operations $U_i$ have the effect of permuting the weightings of projecting into the different subspaces in $\hat{P}^{\prime}_\text{even}$. For example $U_2\hat{P}^{\prime}_\text{even}U_2$ has the same form as $\hat{P}^{\prime}_\text{even}$ with the weightings redistributed according to the relabelling: $1\leftrightarrow2$, $3\leftrightarrow 4$. Expanding out Equation \ref{eqn::Xflip protocol} we find 16 `even' terms $P^{(i)}_{\text{even}}\rho P^{(j)}_{\text{even}}$, 16 `odd' terms $P^{(i)}_{\text{odd}}\rho P^{(j)}_{\text{odd}}$ and 32 `cross' terms $P^{(i)}_{\text{even}}\rho P^{(j)}_{\text{odd}}$. We add another level to our protocol, applying ($\mathbb{1}\mathbb{1}\mathbb{1}\mathbb{1}$) or ($ZZZZ$) with probability $1/2$ to kill off the cross terms.

We then find that it is possible to re-express $P^{\text{smooth}}_{\text{even}}(\rho)$ as the probabilistic sum of perfect odd and even parity projections, followed by $Z$ errors on either one or two data qubits,

\begin{equation}
\label{eqn::kraus}
\begin{aligned}
P^{\text{smooth}}_{\text{even}}(\rho)=\ & \omega_{ \text{even}} P_{ \text{even}}\rho P_{ \text{even}} +\\
&\Gamma_{ \mathbb{1}ZZ\mathbb{1}} P_{ \text{even}}^{\mathbb{1}ZZ\mathbb{1}}\rho P_{ \text{even}}^{\mathbb{1}ZZ\mathbb{1}}+\\
&\Gamma_{ \mathbb{1}Z\mathbb{1}Z} P_{ \text{even}}^{\mathbb{1}Z\mathbb{1}Z}\rho P_{ \text{even}}^{\mathbb{1}Z\mathbb{1}Z}+\\
&\Gamma_{ \mathbb{1}\mathbb{1}ZZ } P_{ \text{even}}^{\mathbb{1}\mathbb{1}ZZ}\rho P_{ \text{even}}^{\mathbb{1}\mathbb{1}ZZ}+\\
&\Delta_{Z\mathbb{1}\mathbb{1}\mathbb{1}} P_{ \text{odd}}^{Z\mathbb{1}\mathbb{1}\mathbb{1}}\rho P_{ \text{odd}}^{Z\mathbb{1}\mathbb{1}\mathbb{1}} +\\
&\Delta_{ \mathbb{1}Z\mathbb{1}\mathbb{1}} P_{ \text{odd}}^{\mathbb{1}Z\mathbb{1}\mathbb{1}}\rho P_{ \text{odd}}^{\mathbb{1}Z\mathbb{1}\mathbb{1}}+\\
&\Delta_{ \mathbb{1}\mathbb{1}Z\mathbb{1}} P_{ \text{odd}}^{\mathbb{1}\mathbb{1}Z\mathbb{1}}\rho P_{ \text{odd}}^{\mathbb{1}\mathbb{1}Z\mathbb{1}}+\\
&\Delta_{ \mathbb{1}\mathbb{1}\mathbb{1}Z} P_{ \text{odd}}^{\mathbb{1}\mathbb{1}\mathbb{1}Z}\rho P_{ \text{odd}}^{\mathbb{1}\mathbb{1}\mathbb{1}Z},
\end{aligned}
\end{equation}
where the Kraus operators \begin{equation}P_{\text{even/odd}}^{U_aU_bU_cU_d}=(U_aU_bU_cU_d) P_{\text{even/odd}},\end{equation} are each applied with a certain probability. Writing Equation \ref {eqn::kraus} in terms of  $P^{(i)}_{\text{even}}\rho P^{(j)}_{\text{even}}$ and $P^{(i)}_{\text{odd}}\rho P^{(j)}_{\text{odd}}$ and equating with Equation \ref{eqn::Xflip protocol}, we see that the probabilities can be expressed as in terms of the weightings $c_i$, $s_i$ as follows,
\begin{equation}
\begin{aligned}
\left(\begin{array}{c}
\omega_{\text{even}} \\
\Gamma_{ \mathbb{1} ZZ \mathbb{1} } \\
\Gamma_{ \mathbb{1} Z\mathbb{1} Z} \\
\Gamma_{ \mathbb{1} \mathbb{1} ZZ} \\
 \end{array}\right)&=\frac{1}{4} \left(\begin{array}{cccc}
1 & 1 & 1 & 1 \\
1 & -1 & -1 & 1 \\
1 & -1 & 1 & -1 \\
1 & 1 & -1 & -1 \\
\end{array}\right)\left(\begin{array}{c}
C_1 \\
 C_2 \\
 C_3 \\
 C_4 \\
\end{array}\right) \\ \\
\left(\begin{array}{c}
\Delta_{ Z \mathbb{1} \mathbb{1} \mathbb{1} } \\
\Delta_{ \mathbb{1} Z \mathbb{1} \mathbb{1} } \\
\Delta_{ \mathbb{1} \mathbb{1} Z \mathbb{1} } \\
\Delta_{ \mathbb{1} \mathbb{1} \mathbb{1} Z} \\
 \end{array}\right)&= \frac{1}{4}\left(\begin{array}{cccc}
1 & 1 & 1 & 1 \\
1 & 1 & -1 & -1 \\
1 & -1 & 1 & -1 \\
1 & -1 & -1 & 1 \\
\end{array}\right)\left(\begin{array}{c}
 S_1 \\
 S_2 \\
 S_3 \\
 S_4 \\
\end{array}\right)
\end{aligned},
\end{equation}
where $C_1 = \frac{1}{4}\left(c_1^2+c_2^2+c_3^2+c_4^2\right)$, $C_2 = \frac{1}{2}\left(c_1 c_2 +c_3 c_4\right)$, $C_3 = \frac{1}{2}\left(c_1 c_3 +c_3 c_4\right)$, $C_4=\frac{1}{2}\left(c_1 c_4 +c_2 c_3\right)$, $S_1 = \frac{1}{4}\left(s_1^2+s_2^2+s_3^2+s_4^2\right)$, $S_2 = \frac{1}{2}\left(s_1 s_2 +s_3 s_4\right)$, $S_3 = \frac{1}{2}\left(s_1 s_3 +s_3 s_4\right)$ and $S_4 = \frac{1}{2}\left(s_1 s_4 +s_2 s_3\right)$.\\

\global\long\def\cosd{\mathcal{C}}
\global\long\def\sind{\mathcal{S}}
\global\long\def\id{\mathbb{1}}

Defining $\cosd_{i}=\cos(\frac{\delta_i}{2})$ and  $\sind_{i}=\sin(\frac{\delta_i}{2})$, the explicit forms of the resulting probabilities expressed as functions of the phase errors $\delta_i$ are

\begin{eqnarray*}
\omega_{\mathrm{even}} & = & \left[\cosd_{1}\cosd_{2}\cosd_{3}\cosd_{4}+\sind_{1}\sind_{2}\sind_{3}\sind_{4}\right]^{2}\\
\Gamma_{\id ZZ\id} & = & \left[\cosd_{1}\sind_{2}\sind_{3}\cosd_{4}+\sind_{1}\cosd_{2}\cosd_{3}\sind_{4}\right]^{2}\\
\Gamma_{\id Z\id Z} & = & \left[\cosd_{1}\sind_{2}\cosd_{3}\sind_{4}+\sind_{1}\cosd_{2}\sind_{3}\cosd_{4}\right]^{2}\\
\Gamma_{\id\id ZZ} & = & \left[\cosd_{1}\cosd_{2}\sind_{3}\sind_{4}+\sind_{1}\sind_{2}\cosd_{3}\cosd_{4}\right]^{2}\\
\Delta_{Z\id\id\id} & = & \left[\sind_{1}\cosd_{2}\cosd_{3}\cosd_{4}+\cosd_{1}\sind_{2}\sind_{3}\sind_{4}\right]^{2}\\
\Delta_{\id Z\id\id} & = & \left[\cosd_{1}\sind_{2}\cosd_{3}\cosd_{4}+\sind_{1}\cosd_{2}\sind_{3}\sind_{4}\right]^{2}\\
\Delta_{\id\id Z\id} & = & \left[\cosd_{1}\cosd_{2}\sind_{3}\cosd_{4}+\sind_{1}\sind_{2}\cosd_{3}\sind_{4}\right]^{2}\\
\Delta_{\id\id\id Z} & = & \left[\cosd_{1}\cosd_{2}\cosd_{3}\sind_{4}+\sind_{1}\sind_{2}\sind_{3}\cosd_{4}\right]^{2}
\end{eqnarray*}

We have thus shown that random application of one of a set of four unitaries before and after an `imperfect' parity projection $\hat{P}^{\prime}_\text{even}$ can be expressed as a superoperator on the data qubits. This has the form of the probabilistic application of `perfect' parity projectors followed by Pauli-Z errors on subsets of the data qubits. When the phase errors $\delta_i$ are small the most probable operation is the desired perfect even parity projection $P_{\text{even}}$ with no errors. This information on stabilizer performance then enables classical simulation of a full planar code array, and its fault tolerance threshold can be assessed. 

The above considers the superoperator for a noisy parity projection in our probabilistic protocol predicated on obtaining the `even' result when measuring the probe. A similar result can be derived in the case that the probe is measured and found in the $\ket{-}$ state.

The erroneous odd parity projection in this case is thus
\begin{equation}
\hat{P}^{\prime}_\text{odd}=  \ S^d_4 S^c_3 S^b_2 S^a_1 - Z^d Z^c Z^b  Z^a S^{'d}_4 S^{'c}_3 S^{'b}_2 S^{'a}_1
\end{equation} 

Again randomly applying our four $U_i$ allows us to derive a superoperator $P^{\text{smooth}}_{\text{odd}}(\rho)$ in terms of perfect parity projections and one- and two-qubit $Z$ errors. This takes the form

\begin{equation}
\begin{aligned}
P^{\text{smooth}}_{\text{odd}}(\rho)=\ & \omega_{ \text{odd}} P_{ \text{odd}}\rho P_{ \text{odd}} +\\
&\lambda_{ \mathbb{1}ZZ\mathbb{1}} P_{ \text{odd}}^{\mathbb{1}ZZ\mathbb{1}}\rho P_{ \text{odd}}^{\mathbb{1}ZZ\mathbb{1}}+\\
&\lambda_{ \mathbb{1}Z\mathbb{1}Z} P_{ \text{odd}}^{\mathbb{1}Z\mathbb{1}Z}\rho P_{ \text{odd}}^{\mathbb{1}Z\mathbb{1}Z}+\\
&\lambda_{ \mathbb{1}\mathbb{1}ZZ } P_{ \text{odd}}^{\mathbb{1}\mathbb{1}ZZ}\rho P_{ \text{odd}}^{\mathbb{1}\mathbb{1}ZZ}+\\
&\zeta_{Z\mathbb{1}\mathbb{1}\mathbb{1}} P_{ \text{even}}^{Z\mathbb{1}\mathbb{1}\mathbb{1}}\rho P_{ \text{even}}^{Z\mathbb{1}\mathbb{1}\mathbb{1}} +\\
&\zeta_{ \mathbb{1}Z\mathbb{1}\mathbb{1}} P_{ \text{even}}^{\mathbb{1}Z\mathbb{1}\mathbb{1}}\rho P_{ \text{even}}^{\mathbb{1}Z\mathbb{1}\mathbb{1}}+\\
&\zeta_{ \mathbb{1}\mathbb{1}Z\mathbb{1}} P_{ \text{even}}^{\mathbb{1}\mathbb{1}Z\mathbb{1}}\rho P_{ \text{even}}^{\mathbb{1}\mathbb{1}Z\mathbb{1}}+\\
&\zeta_{ \mathbb{1}\mathbb{1}\mathbb{1}Z} P_{ \text{even}}^{\mathbb{1}\mathbb{1}\mathbb{1}Z}\rho P_{ \text{even}}^{\mathbb{1}\mathbb{1}\mathbb{1}Z}
\end{aligned}
\end{equation}
\smallskip
where the probabilities are given by
\smallskip

\begin{eqnarray*}
\omega_{\mathrm{odd}} & = & \left[\cosd_{1}\cosd_{2}\cosd_{3}\cosd_{4}-\sind_{1}\sind_{2}\sind_{3}\sind_{4}\right]^{2}\\
\lambda_{\id ZZ\id} & = & \left[\cosd_{1}\sind_{2}\sind_{3}\cosd_{4}-\sind_{1}\cosd_{2}\cosd_{3}\sind_{4}\right]^{2}\\
\lambda_{\id Z\id Z} & = & \left[\cosd_{1}\sind_{2}\cosd_{3}\sind_{4}-\sind_{1}\cosd_{2}\sind_{3}\cosd_{4}\right]^{2}\\
\lambda_{\id\id ZZ} & = & \left[\cosd_{1}\cosd_{2}\sind_{3}\sind_{4}-\sind_{1}\sind_{2}\cosd_{3}\cosd_{4}\right]^{2}\\
\zeta_{Z\id\id\id} & = & \left[\sind_{1}\sind_{2}\sind_{3}\cosd_{4}-\cosd_{1}\cosd_{2}\cosd_{3}\sind_{4}\right]^{2}\\
\zeta_{\id Z\id\id} & = & \left[\cosd_{1}\cosd_{2}\sind_{3}\cosd_{4}-\sind_{1}\sind_{2}\cosd_{3}\sind_{4}\right]^{2}\\
\zeta_{\id\id Z\id} & = & \left[\cosd_{1}\sind_{2}\cosd_{3}\cosd_{4}-\sind_{1}\cosd_{2}\sind_{3}\sind_{4}\right]^{2}\\
\zeta_{\id\id\id Z} & = & \left[\sind_{1}\cosd_{2}\cosd_{3}\cosd_{4}-\cosd_{1}\sind_{2}\sind_{3}\sind_{4}\right]^{2}
\end{eqnarray*}

A similar analysis can be applied to the three-qubit stabilizers which define the boundaries of the planar code. The superoperators for these edge stabilizers are also expressed as perfect $P_{\text{even}}$ and $P_{\text{odd}}$ projections followed with some probability by one- and two-qubit $Z$ errors.
 
\bigskip 

\section{effect of other parameters on the threshold\label{Appendix:furtherthresholds}}

In main paper we determine the threshold value of the required donor implantation accuracy, as this is the crucial parameter for fabricating a large scale device. In doing so we fixed the errors associated with manipulations and measurement of the qubits to values currently achievable in the experimental state-of-the-art. The threshold is in fact a `team effort': if we are able to reduce, for example, the measurement error then greater error in the other parameters can be tolerated. The `thresholds' we determined are thus single points in a vast parameter space.  In this Appendix we investigate further the effect that changing some of these parameters will have on the `donor implantation error' threshold values determined in Fig.~\ref{resultsFig}. Having made our code openly available we hope that the interested reader will find it easy to make further investigations based on their own favoured parameter regimes.

 \begin{figure*}[b]
\centering
\includegraphics[width=\columnwidth]{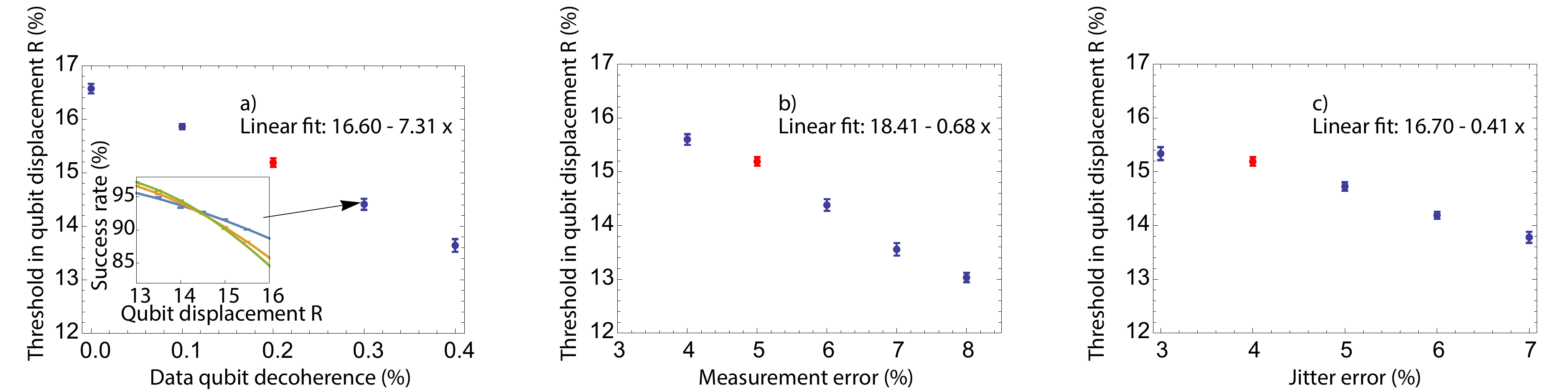}
\caption{\label{Threshold variations} Variation in the qubit displacement threshold depending on a) data qubit decoherence; b) measurement fidelity and c) jitter error. Each data point corresponds to a a full threshold simulation as shown in the insert to a). All data in this figure are for the case of a ``circular probe'' orbit with ``pillbox'' distributed qubit positional errors. As such the red data points correspond to the threshold result shown in Fig.~\ref{resultsFig}(e).}
\label{threshold variation}
\end{figure*}

Fig.~\ref{threshold variation} shows how the threshold value for qubit misplacement depends on other key error parameters. In each graph one error parameter is varied while the others remain fixed to the values stated in Fig.~\ref{Sources of error}(c). The simulations and the resulting fits show that data qubit decoherence is a key source of error to minimise in comparison with measurement and `jitter' errors, at least in this region of the parameter space. Doubling the data qubit decoherence in this regime will have a more deleterious effect on the tolerance of implantation errors than a similar doubling of jitter or measurement error. However, the low rate of decoherence in donor qubits in silicon is one of the great strengths of the system. 

Interestingly, jitter errors (which corresponds to random fluctuations in the strength of the dipole interaction, which one might envisage being due to random timing error) seem to be even more well tolerated than measurement errors, which are generally seen as one of the less crucial sources of error for surface code thresholds. One might imagine that - in a future where donor placement is very far below threshold - characterisation of the device and individual control of probe orbits might mean that the fixed misplacement errors can become random errors in how well we calibrate our device to the misalignments. This would make `jitter' the main source of error, as such it is encouraging to see it is well tolerated.

\color{black}
\section{Considering the `dipolar background'} \label{Appendix::dipole background}

The simple of model of decoherence that we employ in our simulations can be improved to take in to account correlated errors arising from the magnetic dipole-dipole interactions of the data qubits with each other. These will lead to correlated pairs of errors occurring between the qubits, the probability of which decreases with the distance between them.

In this appendix we investigate the more nuanced model of decoherence that incorporates errors of this kind. It has already been shown in~\cite{FowlerCorrelated} that, despite the provable lack of threshold for these kind of errors in the surface code, practically speaking they are tolerable in that they can be suppressed to a desired degree. In fact that paper demonstrates that correlated pairs of errors whose probability decays with the \textit{square} of the distance separating the two qubits are well-handled, this is an even longer range interaction than the one considered here.

\begin{figure*}[t]
\centering
\includegraphics[width=\columnwidth]{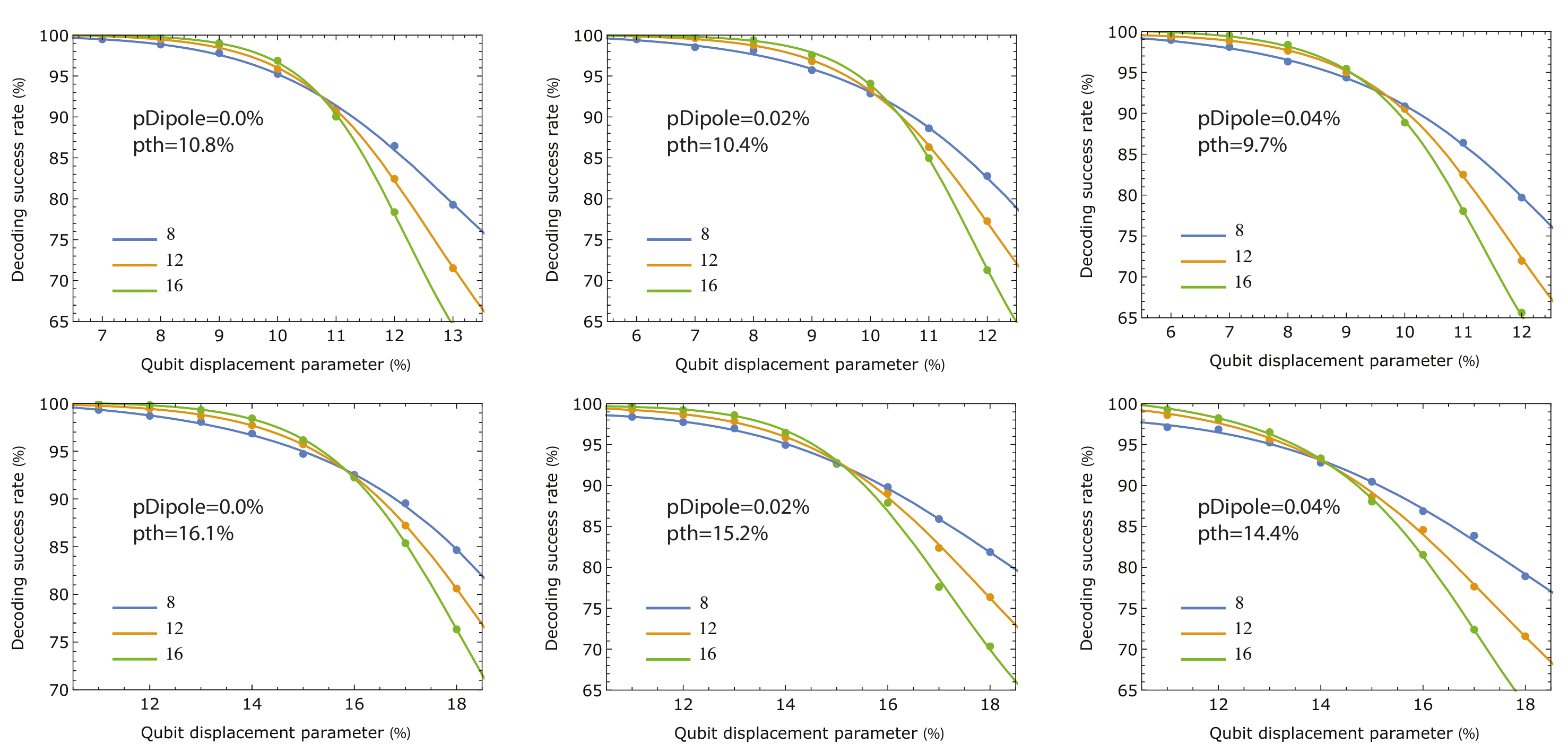}
\caption{\label{dipole background} Variation in the qubit displacement threshold as the dipolar coupling of the data qubits is turned on. All data in this figure are for the case of `pillbox' distributed qubit positional errors, as in Fig.~\ref{resultsFig}(b) and (e). The upper three panels are for an abrupt probe orbit, the lower three for a smooth circular orbit. The other sources of error are fixed to the same values as in the main text of the paper as per Fig.~\ref{Sources of error}, with the exception of data qubit decoherence which we now set to \SI{0.1}{\percent} to reflect than in earlier simulations we had modelled this being due in part to these dipolar couplings.}
\label{fig::dipole background}
\end{figure*}

In Fig.~\ref{fig::dipole background}. we show threshold plots similar to those presented in the main text and Appendix~\ref{Appendix:furtherthresholds} which demonstrate that the position of the threshold in terms of the qubit displacement varies slowly as we turn on correlated errors on nearest- and next-nearest-neighbours in the data qubit array. Here we set data qubit decoherence as modelled previously, that is as IID single qubit Pauli errors, to occur with a small but non-zero level $p=0.001$, so that the new correlated errors are the dominant effect for the data qubits. We leave other sources of error at the same level as in the other reported simulations.  We then apply correlated errors according to the following procedure.

The Hamiltonians of the dipole-dipole interactions between the data qubits is:
\[
H_{ij }=\frac{J}{r^3}\big( \boldsymbol{\sigma}_i\cdot \boldsymbol{\sigma}_j - 3({\boldsymbol{\hat r}}\cdot \boldsymbol{\sigma}_i )({\boldsymbol{\hat r}}\cdot \boldsymbol{\sigma}_j )\big).
\]

We take an approximation that considers only one- and two-nearest neighbours, which comprise a data qubit and its nearest eight counterparts. Depending on the orientation of the pairs and distance separating them their evolution is governed by slightly different Hamiltonians. For example, the Hamiltonians describing a qubit $1$ and its eight pairwise interactions with its nearest and next nearest neighbour neighbours are:

\[
H_{12}=H_{14}=\frac{J}{D^3}\big( -2XX+YY+ZZ \big),
\]

\[
H_{13}=H_{15}=\frac{J}{D^3}\big( XX- 2YY+ZZ \big),
\]
and
\[
H_{16}=H_{17}=H_{18}=H_{19}=\frac{J}{2\sqrt{2}D^3}\big( -\frac{1}{2}XX-\frac{1}{2}YY+ZZ -\frac{3}{2}(XY+YX)\big),
\]
where $D$ is the distance between two nearest neighbour data qubits and the axes and qubit numberings are defined in Fig.~\ref{fig::NNdiag}.

The period of time for which this coupling runs can be estimated as the time to complete a round of stabilizers
\[
t_{stabilizer}=\kappa \  2\pi \frac{d^3}{J}
\]

where $\kappa \in [2,\sim80]$ is a parameter which reflects that in a slow orbit -- such as the smooth circular motion, or a protocol which requires a larger number of orbits per round of stabilizers -- there is a longer time between the completion of full stabilizer rounds. For example, an abrupt motion with 2 orbits required for a full round (i.e. one for $X$ parity measurements and one for $Z$ parity measurtments) corresponds to $\kappa=2$ and slow circular motion with the same number of rounds corresponds to $\kappa=20$.

   \begin{figure*}
\centering
\includegraphics[width=.7\columnwidth]{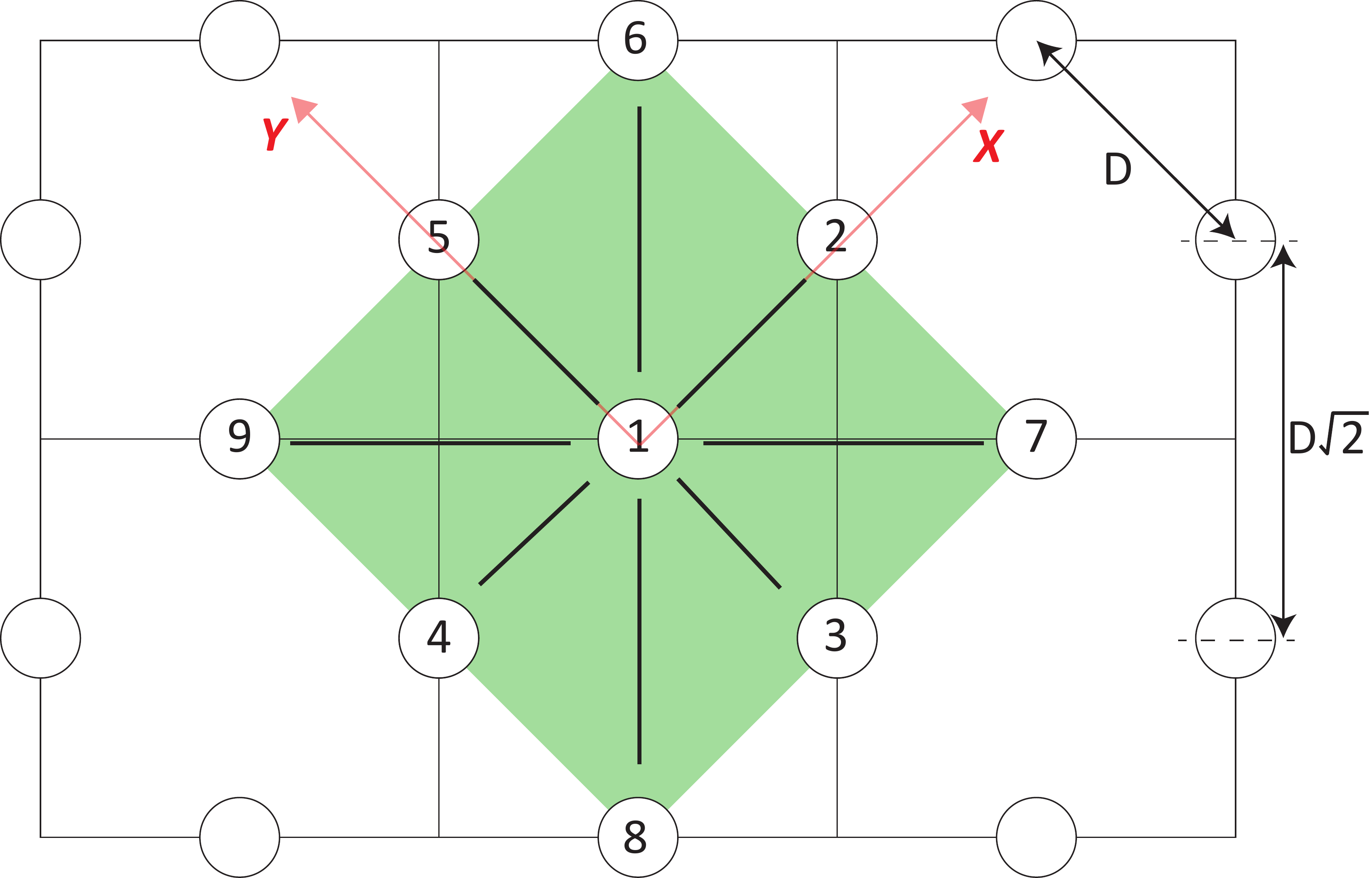}
\caption{\label{fig::NNdiag} A portion of a larger surface code, with nearest and next-nearest interactions labelled as used in our simulated model of the `dipolar background'. To a good approximation the evolution of the data qubits due to their mutual dipolar interactions is modelled by the interaction of nearest and next-nearest neighbour pairs. The distance between nearest neighbours is D, and the axes are chosen such that nearest neighbours lie either on the X- or Y-axis.}
\label{fig::NNdiag}
\end{figure*}

The evolution of the data qubits and its neighbours in a short period of time according to such a Hamiltonian will have the form

\[
U = \text{exp}(-i\left(\sum_{\text{1NN,2NN}} H_{ij} \right) t) \ket{\psi} \approx \left(\prod_{\text{1NN,2NN}} \text{exp}(-i H_{ij} t)\right)\ket{\psi}.\]

Taylor expanding and retaining the leading order terms, i.e. the errors on two-qubits, one obtains
\[\ket{\psi(t_{stabilizer})} \approx \left(\ket{\psi} + \kappa\ 2\pi \left(\frac{d}{D}\right)^3 \left(aX_iX_j \ket{\psi} + b Y_iY_j\ket{\psi} + c Z_iZ_j \ket{\psi} + \ ...\right)\right)
\]
where $a, b, c ...$ are determined by the relative positions of the data qubits with relation to one another.

At each stabilizer measurement the state of the qubit pairs will be projected into one of the states corresponding to either ``no error'' or one of the possible two-qubit errors with probability given by the square of its amplitude. We thus model the dipolar background in a similar way to our other sources of error: we now inject correlated two-qubit errors at the end of each round with the relative probabilities determined by the forms of $H_{ij}$ above. The new parameter in our model is therefore $p_{dip} = \left(\kappa\ 2\pi \right)^2 \left(\frac{d}{D}\right)^6$.

We find that our threshold plots show the familiar behaviour, i.e a well defined crossing point, for the range of $p_{dip}$ considered. To relate the values of of $p_{dip}$ investigated in Fig.~\ref{fig::dipole background} to our proposed device dimensions, consider an abrupt orbit at the dimensions $d=$ \SI{40}{\nano\meter}  and $D=$ \SI{400}{\nano\meter} discussed in the paper. Then  $p_{dip}\sim$ \SI{0.016}{\percent} which is comparable to the rates investigated in the central plots for which we see only small variation in the qubit displacement threshold. In the main text we suggest that for the slower circular orbits (i.e. increasing $\kappa$ by a factor of 10) using the dimensions $d=$ \SI{33}{\nano\meter}  and $D=$ \SI{700}{\nano\meter} which will achieve the same $p_{dip}$ at the cost of a clock cycle approximately 10 times slower. We note that in both the circular and abrupt cases our simulations show that when setting $p_{dip}=$ \SI{0.04}{\percent}, larger than the values relevant to our proposed dimensions, the threshold in terms of qubit misplacement reduces by approximately one tenth, still allowing generous tolerance of the scheme to fabrication error.

The probability of a two-qubit error decreases with $\frac{1}{r^6}$ in this approximation, the results of \cite{FowlerCorrelated} would suggest that practical surface code quantum computation is certainly possible with such a class of error. Although the dipole background can lead to correlated pairs errors occurring in distant parts of a logical qubit, as far as the code is concerned these just appear as two single qubit errors which it tries to correct. The free-running dipolar coupling of the data qubits does not cause the very damaging classes of error such as long chains or large areas suffering error, which could corrupt the logical qubit more easily. Again see~\cite{FowlerCorrelated}.

\end{document}